\documentclass[12pt,preprint]{aastex}
\pdfoutput=1

\usepackage{amsmath, amscd, amssymb, amsthm,color}

\usepackage{graphicx}

\usepackage{hyperref}
\usepackage{ifpdf}
\usepackage{epstopdf}
\usepackage{color}
\def\douter{$\delta_{\mbox{\scriptsize outer}}$}
\def\dsaddle{$\delta_{\mbox{\scriptsize saddle}}$}
\def\dpeak{$\delta_{\mbox{\scriptsize peak}}$}
\def\Ndens{N$_{\mbox{\scriptsize dens}}$}
\def\Nmerge{N$_{\mbox{\scriptsize merge}}$}
\def\HF{\texttt{HF}}
\def\PH{\texttt{PH}}
\def\KD{\texttt{KD}}

\def\kdtree{\it k\rm D-tree\rm}

\shorttitle{Parallel HOP}
\shortauthors{Skory et al.}

\begin{document}


\title{Parallel HOP: A Scalable Halo Finder for Massive Cosmological Data Sets}


\author{Stephen Skory, Matthew J.~Turk, Michael L.~Norman, and Alison L.~Coil}
\affil{Center for Astrophysics and Space Sciences,\\ University of California,
    San Diego, CA 92093}

\email{\{sskory, mjturk, mlnorman, acoil\}@ucsd.edu}


\begin{abstract}
Modern N-body cosmological simulations contain billions ($10^9$) of dark matter particles.
These simulations require hundreds to thousands of gigabytes of memory, and employ hundreds to tens of thousands of processing cores
on many compute nodes.
In order to study the distribution of dark matter in a cosmological simulation,
the dark matter halos must be identified using a halo finder, which establishes the
halo membership of every particle in the simulation.
The resources required for halo finding are similar to the requirements for the simulation itself.
In particular, simulations have become 
too extensive to use commonly-employed halo finders, such that the computational 
requirements to identify halos must now be spread across multiple nodes and 
cores.
Here we present a scalable-parallel halo finding method called Parallel HOP
for large-scale cosmological simulation
data.  Based on the halo finder HOP, it utilizes MPI and domain
decomposition to distribute the halo finding workload across multiple compute
nodes, enabling analysis of much larger datasets than is possible with the strictly
serial or previous parallel implementations of HOP.
We provide a reference implementation of this
method as a part of the toolkit yt, an analysis toolkit for Adaptive Mesh
Refinement (AMR) data that includes complementary analysis modules.
Additionally, we discuss a suite of benchmarks that demonstrate that this method
scales well up to several hundred tasks and datasets in excess of $2000^3$
particles.
The Parallel HOP method and our implementation can be readily applied to
any kind of N-body simulation data and is therefore widely applicable.
\end{abstract}





\section{Introduction}
The overarching goal of a cosmological N-body simulation is to accurately model the hierarchical
distribution of matter that is observed in the universe.
It is computationally efficient to model the mass distribution with collisionless massive particles that
represent dark matter.
The first step in analyzing the distribution of dark matter in a simulation is to locate the dark matter halos,
which are collapsed over-dense regions that host galaxies.
Finding haloes is a computationally difficult task that requires establishing the halo memberships of all the particles
in a simulation.
Similarly to the N-body simulation itself, it is impractical to compare every pair of particles when doing
calculations and approximations must be made for the sake of efficiency.

The field of halo finders for simulated cosmological data is well-populated. All are compromises
between physical accuracy and computational speed. On one end of the spectrum is the
Friends-of-Friends (FOF) halo finder \citep{1985ApJ...292..371D}, which is simple and computationally fast.
FOF builds halos by linking together all particles that are closer than a certain distance from each other.
Most particles will have several linkages, and by recursively following links between particles, a
halo can be identified from the set of inter-linked particles.
However, FOF does not use any physical property of the halo during selection
and is well known to over-connect halos by filamentary bridges \citep{Hut1998}.
In the same article, \citeauthor{Hut1998} present a method called HOP
that improves halo finding by considering the physical density of each particle rather than just locality.
The density of each particle is found by using a smoothing kernel over the neighboring particles,
and halos are built up by recursively linking particles to their densest nearest neighbor,
forming long chains of particles. These chains are then merged to form the final halos.
Halos identified by HOP are far less filamentary than those of FOF, and are more likely
to be physically meaningful.
However, HOP
does not consider gravitational boundedness (kinetic plus potential energy) when assembling halos.
At higher computational expense, \texttt{DENMAX} \citep{1994ApJ...436..467G} and its improvement
\texttt{SKID}\footnote{\texttt{http://www-hpcc.astro.washington.edu/tools/skid.html}}
compute physical density similarly to HOP, and also remove gravitationally unbound
particles from halos. \texttt{VOBOZ} \citep{2005MNRAS.356.1222N} uses Voronoi diagrams \citep{Voronoi}
and Poisson statistics to create halos and eliminate unbound particles.

Many modern halo finders have been designed to identify subhalos within larger
host halos, something that HOP (and therefore our parallelization of it) does not do.
Subhalos are gravitationally-bound objects that reside within or close to larger host halos,
such as satellite galaxies.
This is often a multi-step processes, where host halos are first found with a simple method
(FOF is a popular choice) and then each separately analyzed in greater detail
and at higher computational expense.
This is the workflow employed by \texttt{SUBFIND} \citep{2001MNRAS.328..726S},
\texttt{PSB} \citep{Kim:2006p813} and the method described in \cite{Shaw:2007p878}.
Many methods identify self-bound substructures similarly to the \texttt{DENMAX} recipe such as
the \texttt{BDM} method of \citet{Klypin:1997p684}, the \texttt{IsoDen} method of \citet{Pftitzner:1997p750},
and the method described in \citet{Weller:2005p782}.
Halo finders like \texttt{6DFOF} \citep{Diemand:2006p825} and \texttt{HSF} \citep{Maciejewski:2009p895}
locate substructure by extending the halo finding into the phase space of the particles.
Hierarchical FOF \citep{1999ApJ...516..530K} builds a hierarchy of substructure by running
FOF over a range of linking lengths.
Somewhat similarly, Adaptive FOF \citep{vanKampen:1995p944} finds substructure by automatically
adjusting the linking length based on local density.
Hierarchies of substructure can also be found by basing halos on the hierarchy of grids used in
an Adaptive Mesh Refinement (AMR) simulation
(\texttt{MHF} \citet{Gill:2004p781}, and its replacement \texttt{AHF} \citet{2009ApJS..182..608K}).
An interesting method based on HOP, called \texttt{ADAPTAHOP} \citep{Aubert:2004p933},
is capable of locating substructure and offers some ideas for the future direction of Parallel HOP
(see \S\ref{sec:subhalos}).

All modern cosmological simulation codes are run using the distributed parallel model, in which both the memory and 
workload are handled by multiple processing units on many discrete networked computing nodes.
Increasingly, the simulations produce datasets that can only be analyzed on similarly sized resources
to what is required for the simulation itself.
In our research we use the Eulerian AMR hydrodynamical and cosmological code
ENZO \citep{2004astro.ph..3044O, 2007arXiv0705.1556N}, which is capable of running simulations
with billions ($10^9$) of particles that require terabytes of memory.
Traditionally,
HOP has been distributed with ENZO and used as the primary halo finder for ENZO data.
The publicly available version of HOP\footnote{\texttt{http://cmb.as.arizona.edu/$\sim$eisenste/hop/hop.html}}
(and the version distributed with ENZO,
which adds a wrapper that can read ENZO data)
is not capable of analyzing very large datasets because it is not parallel-capable.
Even if a shared memory computer could contain the data for billions of particles,
the computational time for serially examining
that much data would be immense.
Parallel halo finders, such as
\texttt{AHF} and FOF \citep{2005MNRAS.364.1105S}
would be feasible, but
changing halo analysis platforms would
make direct comparison to previous ENZO/HOP results
difficult, impeding both resolution studies and simulation verification.

Here we present a parallelized version of HOP, called Parallel HOP. It is an update to the HOP method
that can analyze arbitrarily large datasets on as many cores and compute
nodes as needed.
In Section \ref{section:HOP} we describe the original HOP method in detail, as well
as the limitations of two previous parallel versions of HOP.
The general Parallel HOP method is outlined step-by-step in Section \ref{method},
which is applicable to any kind of N-body cosmological data.
Our implementation of Parallel HOP is described in Section \ref{HOPimplementation},
which is easily adaptable to non-ENZO data.
We benchmark our implementation in Section \ref{section:pHOPperf},
and show that it can analyze very large datasets efficiently, and that it scales well with core count.
We discuss future possibilities of improving Parallel HOP in Section \ref{section:discussion},
as well as continuing work on running it concurrently with an ENZO simulation as an
inline halo finder.
\ref{apx:code} provides a guide to understanding the functional pieces that make up the publicly-available
source code of our Parallel HOP implementation.

In this paper, computing `cores' refer to the physical units of a computer that execute
mathematical instructions, and `tasks' refer to the executables that run on the cores.
It is possible to run multiple tasks per core, but in this paper all benchmarks are run
with one task per core.
Therefore, `cores' and `tasks' are used somewhat interchangeably, although they mean
different things.

\section{HOP}\label{section:HOP}

Building halos with the original HOP algorithm requires four stages. The first
stage calculates the density of each particle using a local smoothing
kernel based on the distances to and masses of its nearest neighbors.
The default number of nearest neighbors used for the smoothing kernel is \Ndens\ (=65 by default),
and the kernel always includes the self-same particle.
The density for each particle is normalized with respect to the average density of particles in the simulational volume.
Therefore, the average density found using HOP is not the same as the average density of the universe.

The nearest neighbors are found using a \kdtree\ \citep{355745}, which is a space-partitioning data
structure that allows nearest neighbors to be found much more efficiently than by
brute-force.
A brute-force nearest neighbor search requires $O(N^2)$ inter-particle distance calculations,
where $N$ is the number of particles.
A \kdtree\ reduces the number of operations to $O(N\log N)$, with no loss of accuracy.
A \kdtree\ is built by recursively subdividing a set of data points
(in this case, particle positions) into evenly sized `buckets' that typically contain tens of data points.
Each data point is assigned uniquely to a bucket, and the buckets are geometrically associated with
one another in a tree-like data structure that is referenced when performing
nearest neighbor searches.
A \kdtree\ search is much faster than brute-force because a nearest neighbor search needs only
to compute distances to particles in a few nearby buckets, instead of all the particles
in the entire dataset.

The next step in HOP is to link each particle to its densest nearest neighbor, which is the particle in the set of
neighbors with the greatest density. This link may be to the self-same
particle if it is the densest particle in its set of neighbors.
Long chains of particles are built in the third step by `hopping' from lower to higher density particles, following
the links found in the previous step.
Chains terminate on a particle that is its own densest nearest neighbor.
There is no restriction on the shape of the chains,
which means that there is also no restriction on the shape of the final halos.
Finally, geometrically adjacent chains are merged into final halos according
to merging rules and density threshold cutoffs.

A typical set of cutoff values excludes all particles from halos with densities less than 80 times the average
density of particles in the simulation,
and requires that all final halos have at least one particle with a density 240 times the average.
Changing the cutoff values has the strongest effect on the smallest halos.
Raising the cutoff values reduces the number of small halos identified, while
lower values identify more small halos at the cost of large halos becoming
physically extended and over-linked by filamentary low-density bridges.

Chain merging in HOP attempts to merge density maxima inside of density contours, and is done in five steps.
A diagram showing example chains and merging possibilities is shown in Figure \ref{fig:merging}.
First, particles with density below a user-set threshold \douter\ (=80 by default, as mentioned above) are
excluded from becoming members of any final halo.
Second, the particle with maximal density is found for each chain,
and if that particle's density is above a peak density threshold \dpeak\ (= 3\douter\ by default),
it is considered a proto-halo (but it is still considered a chain for future steps, as well).
Third, for all particles in chains, the \Nmerge\ (=4 by default) nearest neighbors are found.
If a nearest neighbor particle has different chain assignment from another, the average combined density 
$\delta_{\mbox{\scriptsize link}}$ of the two particles
is calculated and added to a lookup table of boundary densities between chains.
Note that the inter-chain links can be unidirectional---the particle in chain A that has
a particle in its set of \Nmerge\ nearest neighbors in chain B,
may not itself be in the \Nmerge\ set of any particle in chain B.
Fourth, if two proto-halos have a boundary density
$\delta_{\mbox{\scriptsize link}}\geq$\dsaddle\ (=2.5\douter\ by default),
they are merged into a combined proto-halo.

The final step which merges sub-\dpeak\ chains to proto-halos is the most subtle.
Sub-\dpeak\ chains are recursively merged to proto-halos by following the densest boundaries
between chains ``downhill'' from proto-halos.
As an example referring to Figure \ref{fig:merging}, 
in the situation where
$\delta_{\mbox{\scriptsize BC}}>\delta_{\mbox{\scriptsize CD}}\geq(\delta_{\mbox{\scriptsize AD}},\delta_{\mbox{\scriptsize DE}})$,
chain D is linked to proto-halo B although it has no direct link to chain B.
This specific situation is unlikely, but it illustrates the recursive nature of the sub-\dpeak\ chain
merging process.
Note that because the boundary densities are established between particles on the fringes of chains,
the boundary densities may be unrelated to the peak density values (especially if the link crosses a density contour line).
It is important to mention here, because it is used in Parallel HOP,
that a proto-halo will always merge with all neighboring proto-halos with which it shares a boundary density greater than or equal to \dsaddle.
However, sub-\dpeak\ chains merge only with one neighboring chain, the chain it shares the greatest
(propagated) boundary density with. Therefore the full, global set of chain-neighbor relationships must be found
before sub-\dpeak\ chains can be merged.
This non-local information requirement is one of the most challenging aspects of the
HOP method addressed by Parallel HOP.

The input and runtime parameters of HOP are discussed in more detail in \S\ref{section:parameters}.

\subsection{Versions of HOP}

The publicly available version of HOP
is actually several executables that are run consecutively, by hand.
Before execution, the end-user must convert their particle data into the native HOP format.
There is a re-packaged version of HOP included with ENZO, called ENZOHOP, that
simplifies the process by joining the components
of HOP together into one executable, and can read ENZO data files natively.
While this enables a straightforward pipeline for ENZO data analysis, ENZOHOP
does not improve on HOP in any other way.

HOP is an efficient method for finding halos, but if run serially, the running
time for analysis of a large dataset can be prohibitive. For example, in our
tests running serial HOP on a dataset with $512^3$ particles, HOP requires on
the order of 10 hours and about 8 gigabytes of shared memory. Extrapolation to a
larger datasets shows that the computation time and the memory requirements
will become prohibitively large.

There are two previous attempts to address the problems of speed and memory usage of original HOP.
OpenMP HOP (\S\ref{OpenMP-HOP}) increases the speed by utilizing multiple processing cores on a single
shared memory node, and MPI HOP (\S\ref{MPI-HOP}) attempts to address
both issues by running on multiple nodes and many cores.

\subsection{OpenMP HOP}\label{OpenMP-HOP}

OpenMP\footnote{\texttt{http://openmp.org/}} is a standardized Application Program Interface (API) that allows a C, C++ or Fortran
program to utilize more than one
processing core on a Symmetric Multi-Processing (SMP) shared memory node.
OpenMP HOP, which is part of the MineBench package \citep{MineBench2006,MineBench2006.2}, parallelizes the
most time and computationally-intestive parts of HOP. At eight processors they find a speedup
of more than five over a single processor run for a range of datasets between approximately $6\times10^5$
and $4\times10^6$ particles
(see Figure 1 of \citeauthor{MineBench2006.2}).

However, this implementation does not address the single node memory limits of HOP because of the SMP requirement of
OpenMP. Despite the improvements to computation time, the algorithm is not
fundamentally changed, and thus does not address the maximum dataset size.

\subsection{MPI HOP}\label{MPI-HOP}

Message Passing Interface (MPI\footnote{\texttt{http://www.mpi-forum.org/}}), is an API that enables
multiple programs (or ``tasks'') running on one or more SMP nodes to
communicate and coordinate with each other, and operate as a single
parallel program. The memory and processing cores available
for the full job are the amalgamation of the capabilities of the SMP nodes used.
MPI HOP, as implemented by \citet{MPIHOP2003}, attempts to address the
limitations of the original HOP implementation by using data-parallelism via
the MPI API. This is done by replacing the original serial \kdtree\ 
with a distributed and parallel
\kdtree.  By design, the distributed \kdtree\ very effectively balances the
workload and memory usage across the nodes.
On a dataset with $\sim5\times10^5$ particles the speedup of MPI HOP is similar
to OpenMP HOP through about 8 processors, but flattens at higher processor
counts, and by 64 processors the speedup is about 20 (see Figures 6 \& 8
from \citeauthor{MPIHOP2003}).

According to the text in \citeauthor{MPIHOP2003}, both the hopping and final group merging are performed in parallel.
A thorough inspection of the MPI HOP source code reveals that this is incorrect.
Instead, the entirety of the particle data are collected on one task and the work is done in serial, identically to
original HOP. This means that MPI HOP is no more capable of analyzing large datasets than the original or OpenMP versions of HOP.

\section{Parallel HOP Method}\label{method}

Here we outline each step of the Parallel HOP process in detail.
Parallel HOP addresses the main shortcomings of the versions of HOP described above.
It distributes both the memory and computational load across as many SMP nodes as required, where
the minimum number is set by the memory requirements. This implementation successfully
distributes the most computationally costly parts of HOP and allows large datasets to be quickly analyzed.

The key to the Parallel HOP method is the local nature of the computational kernel.
As mentioned above, the density of each particle depends only on the distances to and masses of the
\Ndens\ nearest neighbors for that particle.
If the representation of a particle in the computational kernel's memory has its full complement of \Ndens\ nearest
neighbors, its density will be calculated correctly. A corollary of this fact is that if all of the neighbors
of a particle also have their full and correct compliment of nearest neighbors, the densest nearest
neighbor link will be correct for that particular particle.

Parallel HOP uses MPI communication to distribute the work and memory load of
the HOP algorithm across multiple SMP nodes.
The method and implementation (see \S\ref{HOPimplementation}) share some similarities to MPI HOP, but there
are some key differences.
Parallel HOP uses a separate and local \kdtree\ on each task, and never copies the
entire particle dataset onto a single task.
Having separate \kdtree s means that the $O(N\log N)$ operations required for a nearest neighbor search
depends on the number of particles on a task, rather than the entire dataset,
which increases overall efficiency.

\subsection{Step By Step}\label{sec:stepbystep}

As a visual aid, please refer to Figure \ref{fig:logicflow}
for a diagrammatic flowchart of the steps described below.

The first step of Parallel HOP is to spatially subdivide the data into
non-overlapping, space-filling subvolumes.  Although the subvolumes are restricted
to being rectangular prisms, each may be of
different size and shape.
This flexibility can be used to evenly distribute the workload across the computational resource,
which reduces inefficiencies and speeds up the run time (see \S\ref{load-balancing}).
Each MPI task is responsible for exactly one subvolume.

Following the identification and mapping of subvolumes to computational tasks,
the particle data in those subvolumes can be read into memory without duplication.  In
Figure \ref{fig:pHOP02} each subvolume has an unique set of particles within its local
boundaries.

Next, a global lookup table is created that lists subvolumes (or, equivalently tasks) that are geometric neighbors.
Subvolumes may be neighbors via
direct or periodic face, edge or corner contact.
The global lookup table of neighbors, and other lookup tables used below, are a quick-search data object with [key, data] pairs, where the key
is an unique identifier (such as MPI task ID or chain label) and the data are values assigned to that particular key.
In Figure \ref{fig:pHOP02}, the global lookup table of subvolume neighbors links subvolumes A and B together
because they share a face boundary.

The next step is to create and fill a buffer region around each subvolume with
particles.
The goal of filling this ``padding'' region with particles is to
provide the correct set of nearest neighbors for all the particles in the
original, unpadded subvolume, including the particles close to the boundaries.
With the correct set of neighbors, all particles in the original, or ``local''
subvolume will have both their correct density and their correct densest nearest
neighbor.  The padding is calculated from the average inter-particle spacing in the subvolume.
Defining the padding length as $L_p$, we calculate the size of the buffer
region with 

\begin{equation}\label{eqn:padding}
L_p=s\sqrt[3]{\mbox{N}_{\mbox{\scriptsize dens}}V/N},
\end{equation}

where $N$ is the number of particles
in the (unpadded) subvolume, $V$ is the volume of the subvolume, and $s(>1)$ is
a relative safety factor to account for variations in particle density on the
boundaries (see \S\ref{section:safety} for a detailed discussion of $s$).
The cube-root of \Ndens\ is included as a multiplier of the
mean inter-particle spacing to approximate the distance to the most distant nearest neighbor.
In Figure \ref{fig:pHOP02}, $L_p$ is shown to be equal on
all sides of each subvolume for simplicity, but it is usually advantageous to
have different values of $L_p$ for each face (see \S\ref{padding}).  Using
communication and a global lookup table of padding values, particles on the
boundaries of subvolumes are copied to the appropriate neighbor.  In Figure \ref{fig:pHOP02},
particle $\alpha_l$, which is a local particle to subvolume A, has been copied
to the padding in subvolume B as $\alpha_p$.  The situation is reversed for
particle $\beta$.  The particles that reside in the padding constitute the only
instance of persistent particle data duplication in this method.

The padding on each face is a function of task count according to periodic boundary conditions.
When running in serial, the single task already contains in memory all the particle data requried,
and padding is unnecessary.
At two tasks, only two faces for each subvolume touch a different subvolume,
so only two faces need to be padded.
Similarly, at four tasks, four faces need padding.
Only at eight tasks and above will all six faces of each subvolume require padding.

Once the subvolumes have been appropriately padded, the domain of the
halo finding problem has been isolated to only the data present on each task.
Each task then independently calculates the densities and nearest neighbors for
all of its particles, including the padded particles.  With the padding in
place, only the fringe padded particles, and none of the local particles,
have their density calculated incorrectly.

Each task builds chains of particles similarly to original HOP, but with
an additional rule.  There may be no chain link created from a padded particle
to any type of particle. However, chain links may be made from local particles to
any type of particle. This is to prevent link duplication (any link having to do
with a padded
particle would otherwise exist in at least one other subvolume), and also false
linkages as a result of an incorrect set of nearest neighbors,
which only affects padded particles.
With sufficient padding, it is a safe assumption that the densest
nearest neighbor of a local particle, even if that neighbor is a padded particle, is correct.
But there is no such guarantee for a padded particle. This logic is far simpler
than an intra-padding distance cutoff, or some kind of nearest neighbor
correctness test. This process is depicted in Figure \ref{fig:pHOP03}, where
each task has built several locally-labelled chains. Chains that terminate on
the same particle are given the same label, such as $B_1$.  Allowed links are
shown with solid lines, and unallowed ``virtual'' links are shown with dashed
lines.

The next step is an optional, local to each task, premerge of neighboring chains that both have peak densities
$\geq$\dpeak\ (proto-halos) and a boundary density $\delta_{\mbox{\scriptsize link}}\geq$\dsaddle\ between them.
This step is not in the original HOP method or code.
Premerging proto-halos increases the parallelism of the overall chain merging process and reduces
the size of the final global boundary density lookup table (discussed below), which reduces the peak memory usage.
In some cases, enabling this step can reduce the full runtime by nearly 50\%.
For every particle in a proto-halo, their \Nmerge\ neighbors are found.
If any of the neighbors are in a different proto-halo, the average density of the pair of particles is added
to a local lookup table of chain pair boundary densities, if and only if it is the maximum boundary density yet seen for that proto-halo pair.
Neighboring particles that are both padded particles are not used when building the lookup table.
This rule prevents incorrect proto-halo merging and duplication of relationships.
This rule is also followed in the (non-optional) final merging step, described below.
When the search is complete, the local lookup table is used to premerge proto-halos.

Because any two proto-haloes that satisfy the above requirements will be merged with or without
premerging, these mergers can be done
at the local level before the global chain density boundary lookup table is created.
The situation is different for sub-\dpeak\ chains. Because they merge with only one neighboring
chain, the global set of chain relationships needs to be found before they can be merged.

The subdivision of the full volume into subvolumes can result in broken chains if they cross a boundary between subvolumes.
To reconstruct the full chains, tasks communicate to their neighbors the chains that terminate on a particle in the padding.
Put another way, the ``virtual'' links are used to identify chains between two
tasks that must be joined into a single chain.
In Figure \ref{fig:pHOP03}, Task A communicates to Task B that padded particle $\beta_p$ is part of chain A$_2$.
Because Task B has particle $\beta_l$ in chain B$_1$,
Task B links the chains A$_2$ and B$_1$ together.
Chains B$_2$ and A$_3$ are similarly linked. The list of linked chains is collected globally, and particles are
reassigned to new chains built from these connections.
In this fashion, chains that are broken across boundaries are rebuilt without the requirement of collecting the entire
chain particle data on a single task.

Because of the way chains are built and linked across subvolumes, there are particles
that are authoritatively assigned to chains on one task, and exist as unassigned
padded particles on other tasks.
In order to prepare for the final chain merging step, all tasks must be informed
of the chain assignments for their padded particles.
In this step, each task communicates the chain assignment for its particles that
are in the padding of their neighbors. This is based on the set of particles that were
communicated earlier to instantiate the padded population, except only the
subset of particles with chain assignments need to be communicated this time.
For example, in Figure \ref{fig:pHOP03} particle $\alpha_l$ is assigned to the
A$_2$+B$_1$ chain, and this step assigns particle $\alpha_p$ on task B
to the same chain.

The last part of Parallel HOP builds the final halos out of the separate chains.
In this step, both sub-\dpeak\ chains and proto-halos are merged using the same logic as in original HOP.
Even if proto-halos were previously premerged, this step can merge proto-haloes that
have become neighbors after the communication steps above.

As in the premerging step, boundary densities between neighboring chains are added to a
local lookup table.
The separate lookup tables are globally merged using communication to find
the full global collection of chain boundaries that is replicated on all tasks.
This global lookup table of chain boundaries can be quite large, especially if premerging is turned off.
Using the global lookup table of boundaries, chain merging
finishes with precisely the same logic as original HOP, with a few
modifications to take advantage of parallelism.
The parallelism attempts to split up the work of using the global lookup
to make the final halos. However, because the final merging step is a non-local operation,
it cannot be distributed using the established domain decomposition, and much of the
work ends up unavoidably duplicated.

The result is shown in Figure
\ref{fig:pHOP04}, that groups (G$_1$ and G$_2$) may have particles in more than one subvolume.
The padded particles have been discarded and groups are made up of only 
the local particles with no duplication.

Once the halos have been identified independently of subvolume, the statistical properties
of each halo (dark matter mass,
center of mass and maximum radius) are calculated in parallel.

\section{Parallel HOP Implementation}\label{HOPimplementation}

Our implementation of Parallel HOP is written as part of `yt'  \citep{SciPyProceedings_46}.
Yt is an analysis
package for AMR data (ENZO in particular) written in the Python programming language. Yt has an array
of capabilities that complement Parallel HOP, such as parallel computation of radial halo profiles and volumetric projections.
The \kdtree\ used by Parallel HOP is KDTREE 2 \citep{Kennel:2004p978}, which is written in FORTRAN and is loaded into
yt as a module. This is a different and more accurate \kdtree\ implementation than the one original HOP uses.

This kind of mixed-code environment exploits the benefits of each language. Python is an object oriented
language with very powerful built-in functions and data types.
By using a highly-optimized, publicly available FORTRAN \kdtree, the most computationally costly parts of the calculation
can be done using compiled code. This is not to say that Python is slow and inappropriate for large scientific calculations.
There are several features and modules of Python that are exploited in Parallel HOP that use optimized and vectorized compiled
libraries in the background.

\subsection{Input and Runtime Parameters}\label{section:parameters}

There are a wide variety of parameters that control how Parallel HOP runs.
Some are `physical' parameters that directly affect the halos located and have the same function in original HOP and Parallel HOP,
others are `technical' that control how Parallel HOP is run, have no analog in original HOP,
and in some cases can have a small effect on the halos.
Most are user-controlled input parameters, and a few are hard-wired in the source code
that are generally not changed.
The parameters are described in detail below and are summarized in Table \ref{table:parameters}.

\subsubsection{Parallel HOP Parameters}

Of the three density threshold parameters (\douter, \dsaddle, \dpeak), only \douter\ is a user-controlled setting in Parallel HOP.
This is following the recommendation for these parameters given by \citeauthor{Hut1998},
as well as the default settings in the source code of the original HOP implementation.
As discussed in the original method paper (see \S3, and the tests of the threshold parameters in \S2.4),
the halos are most sensitive to the value of \douter, and keeping the other two in a fixed
ratio has the advantage of simplifying the operation of the halo finder.
Lower values of \douter\ increases the number of halos, the total mass in halos and the mass of the largest halos,
and higher values have the opposite effect.

In the original method paper, there are three parameters (\Ndens, \Nmerge, N$_{\mbox{\scriptsize hop}}$)
that control how many neighboring particles are used for different steps.
N$_{\mbox{\scriptsize hop}}$ sets how many neighbors are searched over when building chains.
In \S2.4 of the paper, it is shown that of the three parameters the halos are most sensitive to \Ndens,
with higher values reducing the number of small halos.
In the original source code, the defaults are for N$_{\mbox{\scriptsize hop}}$ to be equal to \Ndens(=65), and \Nmerge=4.
Because Parallel HOP is designed to best replicate the results of original HOP,
these are the settings for Parallel HOP as well.

The \douter=80.0 line in Figure \ref{fig:hmf} shows that the default settings of Parallel HOP
are a good match for the \citet{Tinker:2008p964} $\Delta=300$ halo mass function fitting curve.
Also plotted are the \douter=160.0 and 161.6 mass functions, which shows that as expected,
raising the threshold values reduces the number density of halos.

\subsubsection{Load-balancing}\label{load-balancing}

Parallel HOP in yt employs a sophisticated load-balancing algorithm that allows the use of an arbitrary
number of tasks.
It is constructed similarly to the simple recursive bisection over particle positions method used in a typical \kdtree.
However, instead of cutting a (sub)volume into only two new subvolumes,
this method can cut a (sub)volume into several evenly-populated new subvolumes.
The number of cuts in each round, and the total number of cutting rounds, is taken from the prime factorization of the
number of tasks.
The order of operations is such that the first round of cuts subdivides the full volume into a set of equally-populated
subvolumes, where the number of subvolumes is equal to the greatest prime factor.
The remaining cuts proceed similarly in descending order of the remaining prime factors,
cycling in dimension like a typical \kdtree.
If the number of tasks used is a power of two, this process is identical to a typical \kdtree\ construction.

Ideally, the load-balancing would be accomplished by operating on all the particles in the dataset
as that would provide the most accurately balanced subvolumes.
However, it is unnecessary to operate on the entire dataset.
Testing shows that randomly selecting as little as 0.0003\% of the full population,
and load-balancing on that subset can produce subvolumes with load-imbalances with lower
than a 10\% spread in terms of particles per subvolume.

Therefore, to load balance a dataset, the first step is to read in a random subset of
the particle positions, where the default is a conservative, but still quite small, 3\% of all the particles.
For maximum efficiency, the data is read off disk in parallel, and communicated to one task,
which then load-balances the dataset.
The single task then communicates the boundaries of the derived subvolumes back to all the other tasks,
and the analysis moves forward, which includes reading in the full dataset.

Parallel HOP can also be run with load-balancing turned off, in which case the
subvolumes are all the same size and shape. With load-balancing turned off,
however, the work- and memory-load on tasks can be quite disparate.
If the distribution is excessively uneven, there will be tasks with very little workload
that are forced to wait on the overloaded tasks, which is very wasteful and inefficient,
and runtimes will be high.

\subsubsection{Directional Padding}\label{padding}

The distribution of particles within a subvolume can be highly uneven.
In this scenario, the average inter-particle spacing in the subvolume may not be representative of the
inter-particle spacing near the faces of the subvolume.
In particular, if the particles are sparser near a face than average, the padding distance for that face
can be too small to provide enough padded particles for the density kernel.
In the opposite case, the padding will be too large and there will be unnecessarily duplicated data.

The solution is to have a different padding value for each of the six faces a subvolume,
and to calculate the padding as a function of the inter-particle
spacing using just the particles adjacent to that face.
Unlike in Figure \ref{fig:pHOP02}, with this option turned on the padding $L_p$ will generally not be
the same on all sides.
This option ensures that each face of a subvolume is given a sufficient amount of padding, no more and no less.
In tests, this option has the net effect of reducing the amount of duplicated data without changing the results, 
and speeding up the overall calculation by five to ten percent.

The particles adjacent to each face used for directional padding are selected as follows.
The non-directional padding is found using equation \ref{eqn:padding} on the entire subvolume.
Then, particles that are within that distance from each face are used to re-calculate the inter-particle spacing close to the face,
which in turn gives a different padding value for each face.
The padding safety parameter $s$, described in detail below,
is applied to the padding values to get the final, six-faced directional padding values.

\subsubsection{Padding Safety}\label{section:safety}

It is crucial that all `local' particles in each subvolume have their full and correct set of particles in their smoothing kernel.
Because the actual inter-particle spacing can vary over the boundaries of a subvolume compared to the average,
the calculated padding must be increased by some factor to account for low-density pockets.
The padding safety parameter $s$ has a conservative default value of 1.5, but the user can input a different value.

There is not a single correct value of $s$ because the dynamical range of inter-particle spacing
depends on the specific halos and voids present in a dataset.
Smaller values are desirable because less padding means fewer particles are copied during communication,
and less memory is used on each task.
Larger values than what is required have no effect on results, but will slow down the analysis time and uses more memory.
On some datasets, values of $s<1.5$ will suffice, but on others $s$ may need to be increased.
The default value of $s$ is large enough for all datasets we have tested, 
but it is up to the user to ensure it is large enough by varying $s$ over a range to discover what is sufficient.

To demonstrate the effect of the padding safety parameter on the results,
Parallel HOP is run using 8 cores on a dataset with $128^3$ particles (D128) while varying $s$,
and compared to the single-core results for which padding is unnecessary
(see \S\ref{section:pHOPperf} for details on this dataset).
Curves B, C and D in the upper part of Figure \ref{fig:params} show the mean mass change
in halos cross-matched from a parallel run to the single-core run.
With a small safety factor of 0.01, halos at nearly all mass scales are different,
most notably the 40\% difference for the largest mass bin.
This large difference shows that with a too-small safety factor, the largest halo has been cut nearly in half
by the volume subdivision, and not glued back together.
Increasing the safety factor to 0.5 shows a dramatic improvement with only a small bump at $10^7$ M$_{\odot}$,
and by $s=1.0$ the parallel and single-core halos are identical.

\subsubsection{Chain Premerging}

The optional premerging step represents a trade-off between consistency
and performance.
A consequence of premerging is that the halos produced are sensitive
to the subdivision of the dataset, and are slightly altered when compared to
a non-premerged run.
This is because the premerging step is a local operation, and proto-halos
close to the subvolume boundary do not have information about nearby proto-halos
on adjacent tasks, which alters the eventual make-up of the final halos.
It is possible to fix this flaw, but it would require an expensive global
communication step prior to premerging, which eliminates the advantages of premerging.
This variation only arises when comparing two
runs with different settings;
running the analysis twice with identical settings gives
identical results.

There are three main advantages to premerging.
First, because the operation is local to each task, no communication is necessary and
work can be done simultaneously on all tasks which increases the overall level of parallelism.
Second, it reduces the amount of work required (sometimes dramatically) in the final,
global merging step.
Third, premerging can reduce the size of the final chain merging density boundary
lookup table, which can be a large data object.
Because the global table is replicated on all tasks, the effect of reducing its size
is amplified by the number of tasks.
On some datasets premerging can reduce the full runtime by a factor of two,
and lower the peak memory usage significantly.

A dataset with $1024^3$ particles (D1024) is used to demonstrate the changes in halos
because of premerging (see \S\ref{section:pHOPperf} for details on this dataset).
The lower part of Figure \ref{fig:params} quantifies the differences in halos by comparing
the mass in halos cross-matched between two runs of Parallel HOP with different settings.
Curve E shows that the mean fractional change in halo mass
with and without premerging stays below roughly 0.01\%.
For comparison, curve G shows that a 1\% change in \douter\ has a much stronger effect
on halos than permerging, by roughly two orders of magnitude.
The \douter=160.0 and 161.6 lines in Figure \ref{fig:hmf} show that a 1\% change in \douter\ results
in a virtually indistinguishable difference in the ensemble halo properties.
The premerging differences are two orders of magnitude smaller than 1\% of \douter,
and therefore premerging is an acceptable trade-off for the sake of performance.

Curve F highlights that the premerging differences are a function of subvolume boundaries.
Running with 60 cores produces a different set of subvolume cuts than with 64 cores,
which produces a slightly modified set of halos at the same level of variation as in curve E.

\subsubsection{kD-tree Speedup}

The \kdtree\ used here (KDTREE 2) has a feature that makes a resorted internal copy
of the particle position data that speeds up the nearest neighbor searches by approximately 17\%.
This does not translate into a 17\% speedup for the entirety of Parallel HOP,
but if there is enough memory to contain the duplicated particle data, it is a worthwhile option.
It is turned on by default, but the user may turn it off if memory is constrained.

\subsection{Comparison to Original HOP}

This implementation of Parallel HOP can be directly compared to original HOP by running both versions on one processing core on the identical
dataset with the same set of parameters.
Original HOP is about 40\% faster and uses 20\% less memory than Parallel HOP.
This is because
the original HOP \kdtree\ is faster than KDTREE 2, and also from the the additional steps and data objects created
as part of the parallel machinery that occur even when run in serial.

Parallel HOP generally does not produce completely identical halos to original HOP.
This is because
the \kdtree\ used in original HOP calculates the distances between particles incorrectly by approximately
one part in a million ($10^6$), which affects the smoothing kernel density calculations by up to one part in ten thousand.
This difference in distances and densities is enough to make perfect agreement between halos of original HOP and Parallel HOP impossible.
Curve A in Figure \ref{fig:params} shows the fractional change in mass
in cross-matched halos found by HOP and Parallel HOP on D128.
Although the relative errors in smoothing kernel densities are small, the halo masses change by up to 6\% in this example.
Incorrect distance calculations changes the memberships of the set of nearest neighbors for each particle,
modifying the set of neighbors considered when building the initial chains and chain boundary densities.
For particles at the centers of halos with densities much greater than \dpeak\
(the densest particles in D128 have $\delta\approx 5\times10^5$),
a one part in ten thousand change in density is large in absolute terms,
and can shuffle the relative peak densities of proto-halos, which can then modify
how sub-\dpeak\ chains are merged.

\subsection{Analysis Output}

The results of Parallel HOP can be used and output in several ways within yt. The vital statistics
(mass, center of mass, etc.) of each halo can be output as a text file. The particle data
for each halo can be output into HDF5 files\footnote{\texttt{http://www.hdfgroup.org/HDF5/}},
which allows for detailed post-analysis
or visualization of the halos themselves. Halos in yt are represented as data objects that can be
immediately passed to other analysis toolkits in yt, such as a halo profiler module,
Spectral Energy Distribution (SED) generator (if the simulation contains star particles),
or imaging libraries that can make volumetric
projections or cutting slices through the halo core.

\subsection{Modular Portability of Parallel HOP}

The core of our implementation of Parallel HOP is written in such a way that porting it to another
code base is not difficult.
For each task, its input is simply a subset of the particle data and subvolume
boundaries, and its output are the halo assignments and kernel densities for the particles,
and the halo statistics (center of mass, etc\ldots) for all the halos.
The core implementation is agnostic to how the data are stored on disk, how it is read into memory,
and what will be done with the output.
For example, a simple wrapper as been written that reads in raw binary files
containing particle data and calls Parallel HOP.
We encourage inquiries into adapting Parallel HOP for different kinds of datasets.

\section{Parallel HOP Performance}\label{section:pHOPperf}

In this section the results of a suite of benchmarks of Parallel HOP on three ENZO datasets is presented.
The smallest dataset contains $128^3$ dark matter particles in cube 
$1~\mathrm{h}^{-1}~\mathrm{Mpc}$ on a side, the next larger has $512^3$ particles
in a cube $512 ~\mathrm{h}^{-1}~\mathrm{Mpc}$ on a side, and the largest has $1024^3$ in a cube
$5.6~\mathrm{h}^{-1}~\mathrm{Mpc}$ on a side.  These datasets are referred to as
D128, D512 and D1024, respectively.
D128 and D512 have been evolved to a redshift of z=0, and D1024 to z=6.
Parallel HOP is also benchmarked on a fourth dataset with $2048^3$ particles, but at only one core count,
and it is discussed separately in \S\ref{section:large_dataset}.
All of the benchmarks are run with directional padding turned on.

Except for the largest dataset benchmarked, all of the timings
are taken on the Triton Resource\footnote{\texttt{http://tritonresource.sdsc.edu/}}
at the San Diego Supercomputer Center.
The Triton Resource is a high-performance data-intensive machine made up of 
two clusters of SMP nodes.
The Triton Compute Cluster (TCC) contains 256 8-core nodes with 24 GB of shared memory.
The Petascale Data Analysis Facility (PDAF) contains 28 32-core nodes, 20 with 256 GB of shared memory and
8 with 512 GB of shared memory.
Both clusters have access to the same high-performance disk array and run identical executables.
However, a parallel program may use nodes from only one type of cluster at a time,
never both simultaneously.

There are some important differences between these datasets and benchmarks.
All datasets were analyzed with premerging turned on, and D1024 was also run
with premerging turned off.
D128 and D1024 are unigrid datasets with 32 and 4096 total grid ``tiles'' (hereafter simply grids), respectively.
D512 is a seven-level AMR dataset with 392,865 grids.
There are some extra difficulties associated with AMR datasets with high numbers of grids, such as D512.
In ENZO, a particle's data is stored in the most refined grid that covers the position of that particle.
Therefore, a significant fraction, if not all, of the grids in an AMR dataset contain particle data.
In the D512 dataset, reading the particle data requires accessing the data from
two to three orders of magnitude more grids than the other datasets.
This increases the number of disk operations and slows down the reading of data
when compared to a dataset with equal numbers of particles that is not AMR (i.e. unigrid).
In fact, as shown in Figure \ref{fig:full_timings}, at the same core count, reading the particles for D1024 takes
roughly the same time as D512, even though D1024 has eight times the number of particles and
performs load-balancing.

Another difficulty of AMR datasets comes from the startup- and memory-costs
associated with building the hierarchy of grids and their spatial relationships in yt.
The grid hierarchy is referenced when reading data off disk, and each task must have a
full copy of the hierarchy in memory.
The grid hierarchy for D512 uses approximately
700 megabytes of memory, which is not insignificant when each task must hold in memory a
complete copy.

In datasets of smaller cosmological volumes, the distribution of particles can be
highly uneven, while very large volumes are naturally evenly distributed.
Therefore, D128 and D1024 are run
with load-balancing turned on, and D512 has load-balancing turned off.

\subsection{Timing Benchmarks}\label{s:benchmarks}

In order to measure the speed of various parts of Parallel HOP, a special
command is inserted at various points in the code to record precise time stamps
for each task. The locations of the time stamps are chosen to enclose the different
functional categories of Parallel HOP in order to carefully gauge how the code
scales with respect to core count.

\subsubsection{Full Timings}
In Figure \ref{fig:full_timings} the absolute timing in seconds is shown for
each run. The four distinct parts of Parallel HOP are split into subsections of
the bars. `yt Hierarchy' refers to the startup costs incurred by the
mechanics of yt, which includes reading in the ENZO metadata files and building
the grid hierarchy on each task. 
`Reading Data' covers the time taken to
read the particle data off disk and (if selected) the time taken to perform the
load-balancing.  `Parallel HOP' includes all aspects of Parallel HOP, including
the \kdtree\ searching, group-building and halo statistics computation. `Writing
Data' sums up the time taken to write the halo information to a text file, and
one HDF5 file per task containing the particle data for just the halos.

It is clear that the `yt Hierarchy' step is inconsequential to the full timing of Parallel HOP,
as it is difficult to discern in any of the runs. This shows that the runtime of
this implementation of Parallel HOP is not hindered by the yt framework it relies on.

`Reading Data' displays interesting behavior for each of the timings.
From one to eight cores on D128, it falls, but rises again through 32 cores.
The fall is due to increased parallelism, and the rise is most likely a result of resource competition.
The data for D128 is stored in eight separate files, and by 32 cores there are several tasks
accessing each file simultaneously.
As discussed above, because of the nature of AMR datasets, disk I/O represents a larger fraction
of the full runtime for D512 than the other two, particularly at lower core counts.
But the extra challenge of AMR is well parallelized when using more cores;
from 2 to 64 cores the time drops by nearly a factor of 20.
For D1024 between 8 and 256 cores (also a 32 times increase in core count),
the drop is only slightly better than a factor of 4.
D1024 exhibits smaller gains because of the extra cost of the initial round of data reading for load-balancing.

The `Parallel HOP' step represents the bulk of the time for most runs, and is discussed in detail below.

Finally, `Writing Data' is indiscernible in most runs and does not impact runtimes significantly.
Each task is independently writing data to disk, and the amount of data
is much smaller than was read in originally because only the data for particles in halos
are recorded as output.

\subsubsection{Parallel HOP Timings}
In Figure \ref{fig:pHOP_timings} the parts of Parallel HOP algorithm are examined in detail.
Each of the timing blocks plotted is a sum of the maximum time taken by multiple sub-timings in Parallel HOP that fit under
the timing block description.
A sub-timing contributes time to only one of the timing block categories.
Measuring the maximum time in this fashion is more representative of the real computational costs than an average over all tasks
because it exposes inefficiencies more clearly.
Because Parallel HOP is an asynchronous application, the summed maximum timings includes
overlapping processes, which effectively means some time is double-counted.
This is why the full sum of the timings in Figure \ref{fig:pHOP_timings} can
be longer than the corresponding timing in Figure \ref{fig:full_timings}.

`\kdtree\ Searching' covers
the time taken to perform the multiple nearest neighbor searches using KDTREE 2.
The `MPI (+Related)' step covers both MPI communication
activities and operations undertaken by all tasks necessary for parallelism.
`Halo Creation' is the time required
to perform the merging of chains into final groups.
If premerging is turned on, the time taken during that optional operation is included in `\kdtree\ Searching'
(because searching the \kdtree\ constitutes the bulk of the time for premerging),
Finally, `Halo Statistics' includes calculation of the 
halo centers of masses, total masses, maximum radii and bulk velocities.

The timings for the D128 dataset show that it is too small to be efficiently analyzed by Parallel HOP.
Between 1 and 32 cores there is very little increase in speed.
The MPI overhead starts to dominate the process with task count, and
the \kdtree\ timings do not fall quickly enough to compensate for it.
This is because at high core counts, the volume of padding needed for
each task begins to approach the volume of the subvolumes themselves.
This \kdtree\ `workload floor' effect is clear when the locations of the linear scaling triangles
are compared across all four sets of benchmarks.
In all but the D128 timing, the triangles roughly track the top of the \kdtree\ timing sections.
which shows that the \kdtree\ workload scales fairly linearly, except in D128.

The D512 dataset shows much better scaling, except for some odd results in the MPI timings
at 4, 16 and 64 cores.
A close inspection of the output logs shows that because of the particular distribution of particles,
subdividing the full volume without load-balancing turned on at these core counts produces subvolumes that are not as evenly
populated as the other core counts.
The net effect is that tasks wait on other tasks more, which is reflected in the MPI timings.
To a smaller extent it can also be seen in the \kdtree\ timings,
which don't fall as a fast as might be expected at these particular core counts.
Re-benchmarking D512 with load-balancing turned on shows that
the MPI overhead is reduced, but the overall runtimes are increased by
the extra disk I/O required.

The differences between the two D1024 timings highlights the higher level of parallelism provided by premerging chains.
At eight cores, running without premerging is actually faster because there is one fewer
nearest neighbor search over the particles.
But at higher core counts, without premerging the `Halo Creation' step does not shrink as
quickly, and by 256 cores, the premerging run is 25\% faster than the run without.

\subsubsection{Speedup Plots}

The overall and Parallel HOP-only speedups for each dataset are shown in Figure \ref{fig:pHOP_speedup}.
The Parallel HOP-only line excludes any disk-related operations.
At 32 cores, D128 is only 3.1 times faster than a single core run, which shows that Parallel HOP is not appropriate
for small datasets.
D512 shows much better scaling because it is a larger dataset and therefore more appropriate
for Parallel HOP.
Again, the odd behavior at 16 and 64 cores can be seen with depressed points
at those core counts.
The overall speedup for D512 is better than the computational speedup at 64 cores for this same reason,
and also because the disk I/O shows very good parallelism between 2 and 64 cores.
The advantage of premerging is clear between the two D1024 Figures.
At higher core counts the speedup for both premerging curves is significantly higher
than the curves without premerging.
The gap in the two speedup curves for D1024 is primarily attributable to the cost of load-balancing.

The speedup curves depart from linear scaling for many of the same reasons.
Communication inefficiencies and quantities increase with task count,
as does resource competition, which affects D128 and the 256-core D1024 timings
the most strongly (see \S\ref{sec:corespernode}).
The total amount of particles contained in padded regions also rises with
core count, which further depresses performance.
Variations in the density of particles can have an effect on the load-balacing,
both through their bulk distribution as seen in D512 at 16 and 64 cores,
and also on the scale of padding, which contributes to load-imbalances
between tasks, even when load-balancing is turned on.
The workload floor seen in the D128 timings shows that there is a practical ceiling
for the number of tasks for a given dataset.

\subsubsection{Weak-Scaling}\label{sec:weak}

Weak-scaling is a good measure of the overhead in a parallel program incurred by the parallel machinery.
The problem size and the number of cores used are kept in a constant ratio over a range of core counts,
so any increase in total runtime should come from inefficiencies in parallel communication and associated functions.
Another motivation for weak-scaling tests is to model the behavior of a code as problem size increases,
in terms of both particle and core count.
Ideal weak-scaling is flat---the increase in problem size is perfectly compensated for by
the commensurate increase in the number of cores.

By randomly subsampling particles from D1024 and analyzing a range of dataset sizes,
it is possible to perform a weak-scaling study of Parallel HOP.
Because the random subsampling requires modifying the dataset prior to running Parallel HOP,
it is impossible read in the data in precisely the same fashion as the other benchmarks.
An intermediate step stores the subsampled data on disk, which is read in again for analysis
using a different method than the other benchmarks.
Therefore, only the timings for the computational parts of Parallel HOP are plotted in Figure \ref{fig:pHOP_weak}.
The number of particles stored in memory per task is kept constant at roughly $4.2\times10^6$,
but there is some small variability as a result of imperfect load-balancing.
The maximum number of particles stored on each task is further increased by 10--20\% after padding is considered.

From one core to 256, the computational time required rises from a little over 300 seconds to just under 900.
There are a variety of reasons for this rise, but the primary one is that the workload is nonlinear with total particle count.
As the total number of particles rises, the sampling of the full dataset's nonlinear structure improves,
and this is reflected in the workload-related statistics over the range of benchmarks.
At one core, 8.4\% of all particles are in halos, and there are 3,007 initial chains that merge into 1,467 final halos.
For 256 cores the figures are 11.4\%, 1,170,042, and 124,660, respectively,
which averages to 4,570 initial chains per task.
The 35\% rise in the fraction of particles in halos leads to a 52\% rise in the number of chains per task,
and a 389 times increase in the total number of initial chains.
Some of this increase is caused by chains broken by the subvolume boundaries (and not entirely due to added structure),
which is a function of the number of cores used rather than just particle count.

Each step in Figure \ref{fig:pHOP_weak} is sensitive to these increases.
Increasing the fraction of particles in chains means the \kdtree\ must perform relatively more nearest
neighbor searches for the chain merging steps.
The rise in the fraction of particles in chains and halos, and having more chains globally, means more data must
be communicated and processed during the MPI steps.
More chain interactions via the boundary density lookup table complicates the work of merging chains, which slows
down the Halo Creation step.
Finally, although the calculation of halo statistical properties is parallelized,
and the number of halos rises slower than cores (85 times from one to 256 cores),
because halos are global objects, the work cannot be perfectly spread across the cores.

There are some secondary reasons for the rise in runtime.
As mentioned in \S\ref{sec:stepbystep}, between one and eight tasks the number of padded
faces for a subvolume rises from zero to six, which in turn increases the workload for the \kdtree\ over that range.
With the increased number of initial chains, inevitably the load-balance between tasks worsens,
leading to increased communication-related inefficiencies, which is reflected in the MPI timings.

Although the rise in runtime is not ideal, weak-scaling can never perfectly flat in practice.
Indeed, in light of the nonlinear workload increase, the weak-scaling results are very good.
A 256 times increase in total problem size, and all of the nonlinear effects,
merely triples the runtime over the base single-core mark.

It is possible to conceive a highly artificial dataset that would reduce the nonlinear effects as follows.
Equally-sized clumps of particles evenly gridded in space are placed such that each task contains in memory one clump at the center of its subvolume.
No clumps are close to a subvolume boundary, so no chains need reconnecting
and padding is unnecessary, nor do any chains merge with chains from a different subvolume.
In this scenario, it is expected that the weak-scaling would be much closer to
ideal because nearly all the work has been completely localized.
However, this is not how datasets are created in practice, and the weak-scaling results
presented here are closer to a typical workflow.
It is often beneficial to perform low-resolution survey simulations that help tune the
settings used for a large and expensive production simulation.
When scaling up from the survey simulations to the full size, the behavior of Parallel HOP
on the datasets is much closer to the results shown here than an artificial dataset lacking
the nonlinear workload increases.

\subsection{Memory Usage}

During the benchmark runs, the memory used for each task iss sampled at a number of points during
execution, and the peak level recorded.
The memory usage for the yt mechanisms separate from Parallel HOP are subtracted, leaving
just the memory usage for Parallel HOP. Across all three datasets and core counts there is a nearly universal
ratio of peak memory to number of particles (including padding) of 1 megabyte per 4,500 particles.
Individual tasks ranged by about $\pm$20\% from this ratio.

At first look, this may appear to be a significant amount memory per particle. It must be noted, however, that the peak memory
includes not only the padded particle data, but also the \kdtree\ structure, which is significant, as well
as the particle chain data (such as particle density and densest nearest neighbor link),
and the data objects created for communication.
Data sent during communication exists on the sender and receiver simultaneously, briefly duplicating large data objects,
and it is during communication steps that peak memory levels are reached.

\subsection{Cores Per Node}\label{sec:corespernode}

Most of the benchmark runs use multiple cores per node (see Table \ref{table:corespernode}).
When using more than one core on a SMP node, the cores compete for access to shared computing resources
(such as the memory bandwidth or disk access),
and overall runtimes are increased for the individual tasks when compared to single-task timings.
In isolated tests for this effect we see that for nearly all the benchmarks there is at worst a roughly 10\% slowdown
for the most computationally intensive steps due to inter-core competition.
However, when running on all 32 cores on a Triton PDAF node,
there is a 40\% slowdown over a single task which affects the \kdtree\ operations most strongly.
This only affects the timings of the 256-core D1024 runs, which are the sole
case that use all 32 cores on the PDAF nodes.

\subsection{A Large Dataset}\label{section:large_dataset}

To demonstrate its ability to scale up to larger datasets, Parallel HOP is run on a redshift=0 dataset with 2048$^3$ particles
in a cube $442~\mathrm{h}^{-1}~\mathrm{Mpc}$ per side.
It is run on the Cray XT5 Kraken\footnote{\texttt{http://www.nics.tennessee.edu/computing-resources/kraken}}
at the National Institute for Computational Sciences using 350 cores on
175 nodes, which collectively provide 2.8 terabytes of memory.
The number of nodes is set by the minimum memory requirements needed for the dataset,
which are determined by the memory usage scaling factor.
Load-balancing, premerging and directional padding are turned on.

The full run takes one hour and two minutes. It spends a negligible amount of time
in the `yt Hierarchy' step, 13 minutes in the `Reading Data' step,
41 minutes in `Parallel HOP', and 8 minutes in the `Writing Data' step.
It is expected that the scaling performance of this dataset looks something like
that of D1024 (with premerging) because load-balancing would have the same
effect on the overall performance with core count.

Running Parallel HOP on a dataset this large shows that Parallel HOP is capable
of handling very large datasets efficiently.
Indeed, no previous public implementation of HOP is at all capable of analyzing a dataset with this many particles.
Even if a machine existed that had enough shared memory to contain all the particle data,
the run time would be prohibitively long.

\section{Discussion}\label{section:discussion}

We show here that Parallel HOP is a successful parallelization of the HOP algorithm.
It can analyze datasets much larger than any previous implementation.
It can run on as few as one core, or as many cores as determined
by the memory or running time requirements.

Because Parallel HOP is parallelized using domain decomposition, it is particularly
well suited for use as an inline halo finder in grid-based cosmological
simulation codes.
With inline halo finding, galaxy-related
objects can be introduced as the simulation runs, such as black holes at the centers of large halos, or
galcons \citep{2008ApJ...683L.111A, Arieli:2010p645}, which are a physically-extended
analytical model of galactic properties.
A detailed merger history can be created for all the halos, which can be used to model
the morphologies and star formation histories for the galaxies they contain.
Tests of a preliminary inline implementation have shown this to be an effective
method for both unigrid and AMR ENZO simulations.
The subvolumes on which each task of Parallel HOP runs are identical
to that of the simulation, which rules out load-balancing within Parallel HOP as an option.
However, because Parallel HOP accesses the particle data directly in memory, it eliminates
the sometimes costly `Reading Data' step and speeds up the overall runtime.
Additionally, any toolkit available within yt can be applied at the same time
to the halos and the entire dataset in general,
which enables very powerful and flexible inline analysis.

With regards to the method, the tightest bottlenecks have to do with non-local operations,
such as the `Halo Creation' step.
Although the premerging step greatly alleviates that specific problem,
it doesn't fix the root cause of why `Halo Creation' is slow and unparallelized.
A true fix would involve a careful modification of how the final halos
are created that would remove the global boundary density lookup table,
and replace it by many separate lookup tables synthesized from each of
the initial local lookup tables.

Future improvements should also include optimization of the code for
speed and memory efficiency. Although the most computationally
intensive parts of Parallel HOP in yt already use compiled code,
replacing some or all of the Python with compiled modules would
provide some speedup and memory savings over the current implementation.
Hybrid parallelism, in which both MPI and OpenMP are used simultaneously, is an
effective way to take advantage of modern multi-core processors.
Because Parallel HOP is domain decomposed, hybrid parallelism is
a very plausible way to speed up certain parts of the implementation.
Replacing the FORTRAN \kdtree\ with an implementation that runs
on very fast graphics hardware could provide substantial
speed gains on workstations or clusters with that kind of hardware \citep{1409079}.

It is straightforward to adapt the Parallel HOP method, and indeed
our implementation as well, for other kinds of cosmological simulational data.
Although yt is primarily written with AMR data, because of its modular nature
it is possible to write a new data reader within yt to handle other types of particle data.
Alternatively, the Parallel HOP implementation source code could be adapted
for use inside of a different analysis package fairly easily. Because yt and
Parallel HOP are open-source, anyone may freely copy and modify
the source code for their own needs.

\subsection{Parallel HOP and Subhalos}\label{sec:subhalos}

\begin{figure}[htbp] 
   \centering
   \epsscale{.45}
   \plotone{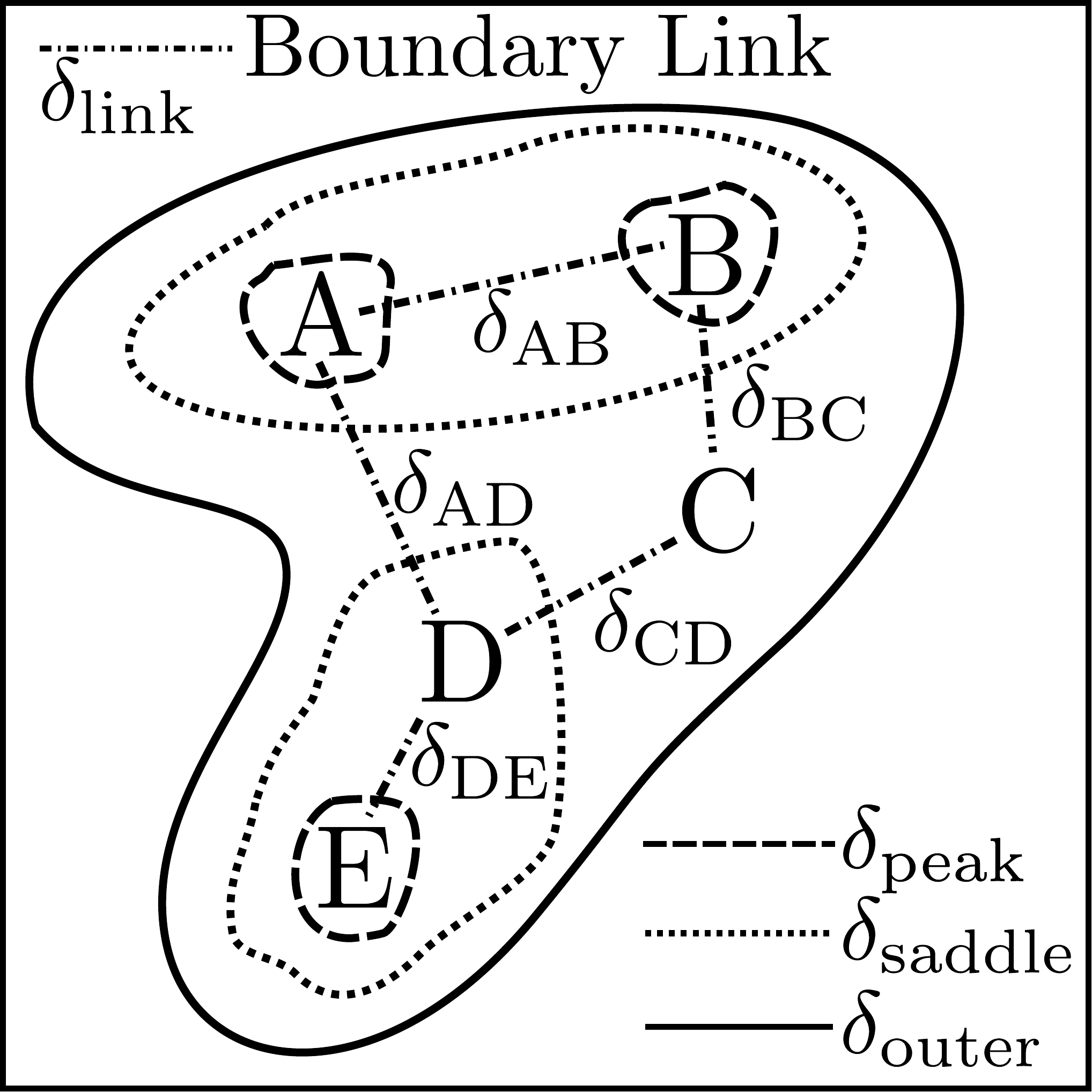}
   \caption{Five particle chains are labeled (A through E) at the location of their peak density particle.
   Density contours surrounding particle chains are shown with lines: \dpeak\ (long-dashed) $>$ \dsaddle\ (short-dashed) $>$ \douter\ (solid).
   The peak of chains A, B, and E lie inside of a \dpeak\ density contour, therefore the chains are considered proto-halos.
   Spatially proximate chains with neighboring particles are shown linked (dot-dashed lines),
   and each link has a boundary density $\delta_{\mbox{\scriptsize link}}$ value
   recorded.
   Proto-halos A and B are merged because their boundary density link lies within a \dsaddle\ density contour
   ($\delta_{\mbox{\scriptsize AB}}$ $\geq$ \dsaddle).
   Although $\delta_{\mbox{\scriptsize DE}}$ $\geq$ \dsaddle, chains D and E do not necessarily merge because D is not
   a proto-halo.
   Chains C and D may merge separately or together with proto-halos A, B or E depending on the relative values
   of the boundary densities $\delta_{\mbox{\scriptsize AD}}$, $\delta_{\mbox{\scriptsize BC}}$, $\delta_{\mbox{\scriptsize CD}}$ and
   $\delta_{\mbox{\scriptsize DE}}$.}
   \label{fig:merging}
\end{figure}

The ability to locate subhalos by using the modifications of HOP in \texttt{ADAPTAHOP}
may be a worthwhile feature to add to Parallel HOP.
Briefly summarized, \texttt{ADAPTAHOP} finds (sub)halos as follows:
Particle densities, and chains of particles (called ``leaves'' by \citeauthor{Aubert:2004p933}),
are found identically to HOP.
Instead of merging chains and proto-halos by all maximum boundary link values,
the ``symmetric'' boundary densities between chains are found and stored in a lookup table.
A ``symmetric'' boundary means that both chains have links to the other,
which is not a requirement in HOP.
Subhalos are identified by adjusting a density threshold parameter $\delta_s$ which is compared to
density boundary links ($\delta_{\mbox{\scriptsize link}}$ from Fig. \ref{fig:merging})
between chains.
All links $\delta_{\mbox{\scriptsize link}} < \delta_s$ are destroyed,
which modifies the chain relationships and selects substructure during final merging,
which is done very similarly to HOP.
For example, if $\delta_s > (\delta_{\mbox{\scriptsize AD}}, \delta_{\mbox{\scriptsize BC}},
\delta_{\mbox{\scriptsize CD}})$ and $\delta_s <= (\delta_{\mbox{\scriptsize AB}}, \delta_{\mbox{\scriptsize DE}})$,
chains A and B, and E and D would form two subhalos.
Chain C may be marked as substructure, depending on whether it has particles with density greater than $\delta_s$.
Finally, subhalos are rejected if their statistical significance falls below the level of Poisson noise.

Because \texttt{ADAPTAHOP} is based on HOP, it should be possible to apply these
modifications to Parallel HOP.
The modifications would be turned on and off depending on user-controlled parameters.
The biggest difference would come from the global chain density boundary lookup table,
which would be constructed differently.
The modifications would also preclude premerging, as the entire set of density boundary links needs to be
preserved for comparison against $\delta_s$.
The output of Parallel HOP would have to be modified to account for the hierarchies of halos and subhalos,
depending on the number of different values of $\delta_s$ input by the user.
The performance of Parallel HOP would be affected by the modifications.
The runtime would increase with the number of link threshold parameters $\delta_s$.
Roughly speaking, the length of the  ``Halo Creation'' step in Fig. \ref{fig:pHOP_timings} would increase linearly.
Also, because particles would be multiply-assigned to (sub)halos, and there would be many more (sub)halos,
the memory requirements would rise, as would the time of the `Halo Statistics' step.

\appendix

\newcommand{\appsection}[1]{\let\oldthesection\thesection
  \renewcommand{\thesection}{Appendix \oldthesection}
  \section{#1}\let\thesection\oldthesection}

\appsection{Parallel HOP Implementation Source Code}\label{apx:code}

This appendix provides a road-map for understanding our Parallel HOP implementation in detail.
The source code is freely available online at the yt homepage\footnote{\texttt{http://yt.enzotools.org/}}
as part of the yt distribution.
Yt has been built and installed on a wide variety of operating systems, from laptops to supercomputers.
It is open-source, and relies on only open-source libraries, so it is free for all to use wherever and however they wish.
Yt has an active developer community, and there is a public mailing list linked from the homepage
that a user can sign up for to receive assistance on technical issues.

\begin{figure}[htbp] 
   \centering
   \epsscale{.8}
   \plotone{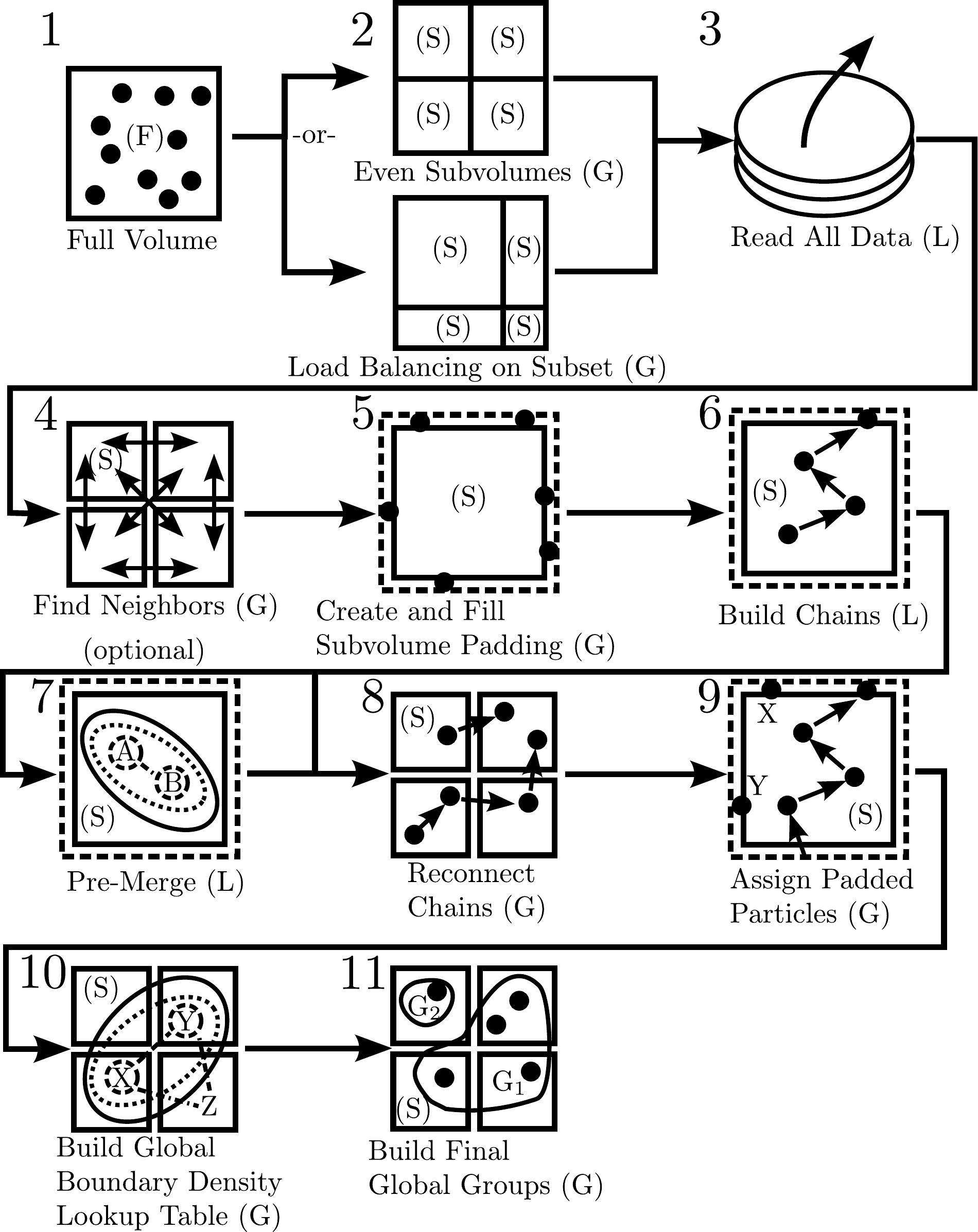} 
   \caption{Step by step flowchart of Parallel HOP.
   Inside the boxes (F) and (S) mean the enclosed volume is the full volume or a subvolume, respectively.
   Chains are labeled with capital letters (e.g. A, B, X), and final groups are labeled by G$_1$ and G$_2$.
   The letters (G) and (L) following the captions label that step as either a global (using MPI communication) or local operation, respectively.
   The numbers next to each glyph refer to the steps and code functions described in \ref{apx:code}.}
   \label{fig:logicflow}
\end{figure}

\begin{figure}[htbp] 
   \centering
   \epsscale{0.6}
   \plotone{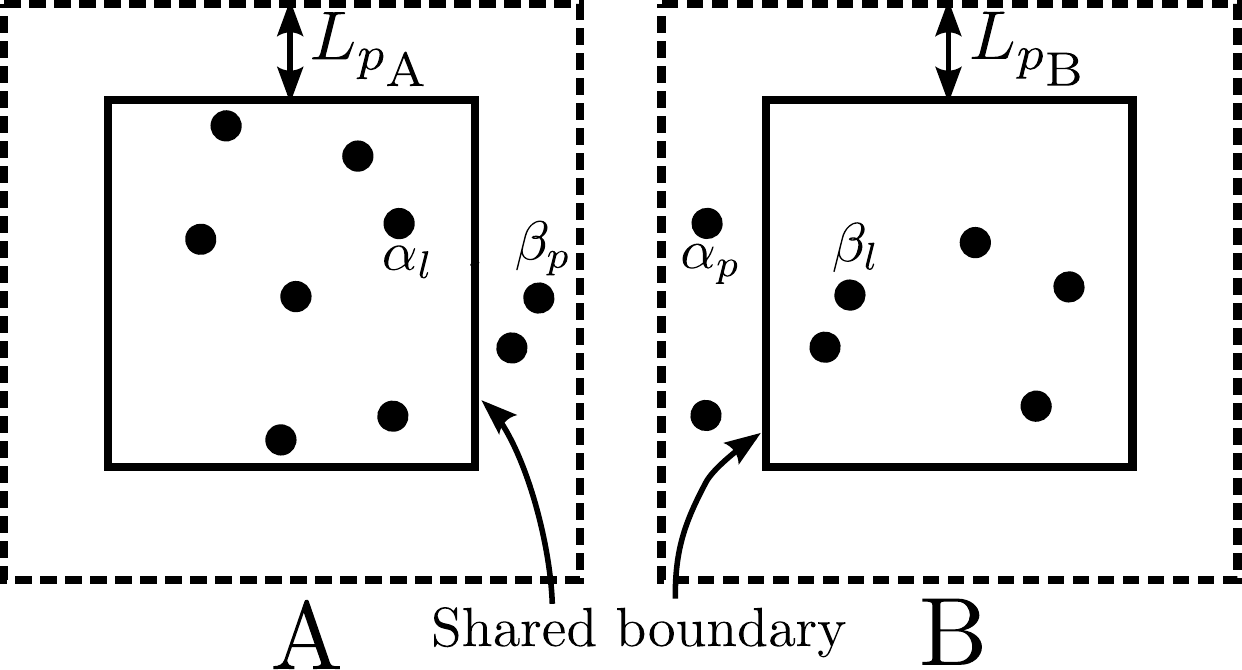}
   \caption{Two subvolumes, A and B, are filled with particles inside their initial boundaries (solid lines).
   A and B share a face-boundary, as shown.
   Each subvolume is surrounded by padding (dashed lines, padding size $L_p$) in which duplicated, padded particles are
   placed, that are copied from neighboring subvolumes using MPI communication.
   Particles $\alpha_l$ and $\beta_l$ are the original \em local \rm copies of their \em padded \rm duplicates
   $\alpha_p$ and $\beta_p$, which are stored in the neighboring padding region.
   Particles $\alpha_p$ and $\beta_p$ are deleted when no longer needed.}
   \label{fig:pHOP02}
\end{figure}

\begin{figure}[htbp] 
   \centering
   \epsscale{0.6}
   \plotone{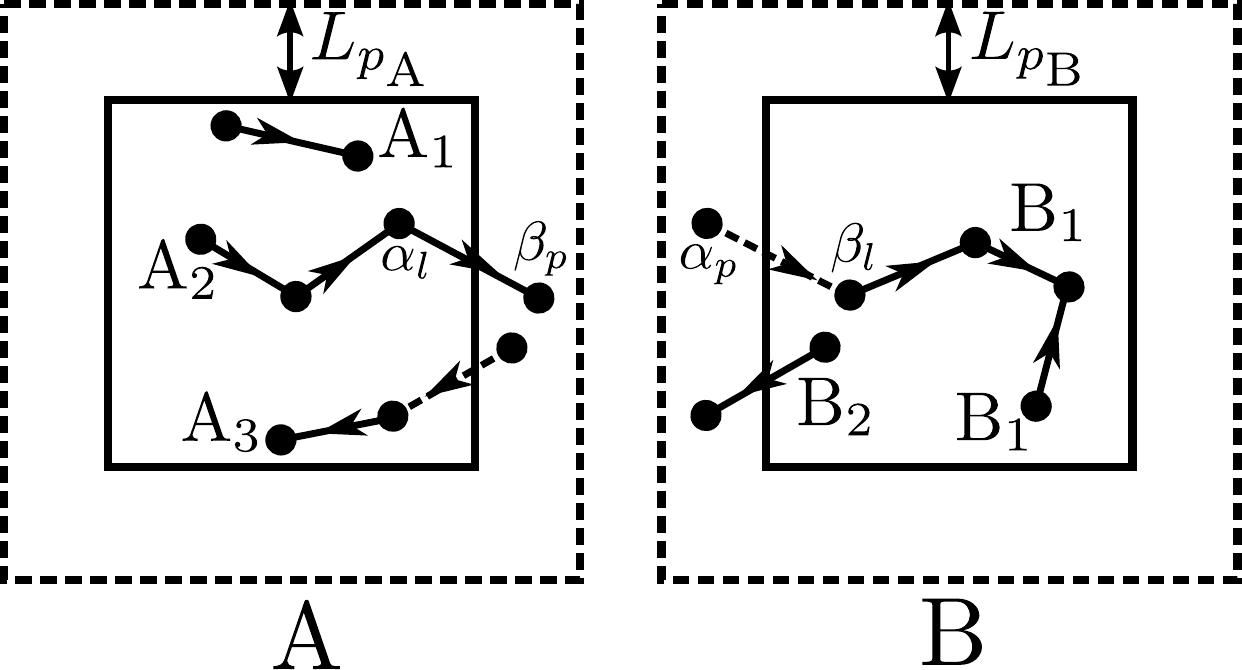}
   \caption{Chains of particles are built in each subvolume according to the linking rules (see text). `Virtual' links that
   are made using communication are shown with dashed lines.}
   \label{fig:pHOP03}
\end{figure}

\begin{figure}[htbp] 
   \centering
   \epsscale{0.6}
   \plotone{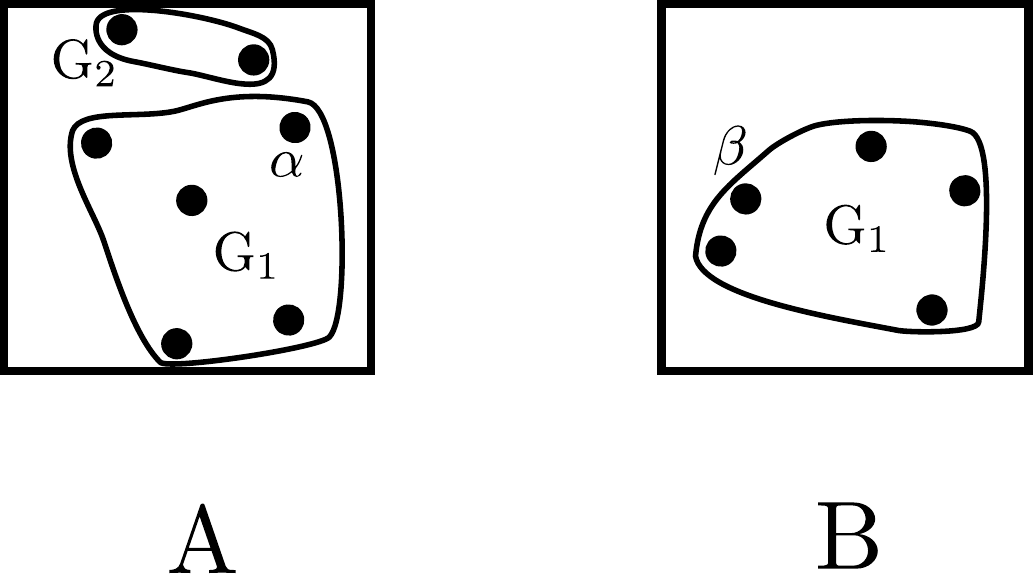}
   \caption{Padded particles are removed and groups (or halos) are formed in each subvolume with global labels. G$_2$ is wholly contained
   in subvolume A and G$_1$ has particles in both.}
   \label{fig:pHOP04}
\end{figure}

\begin{figure}[htbp] 
   \centering
   \epsscale{.7}
   \plotone{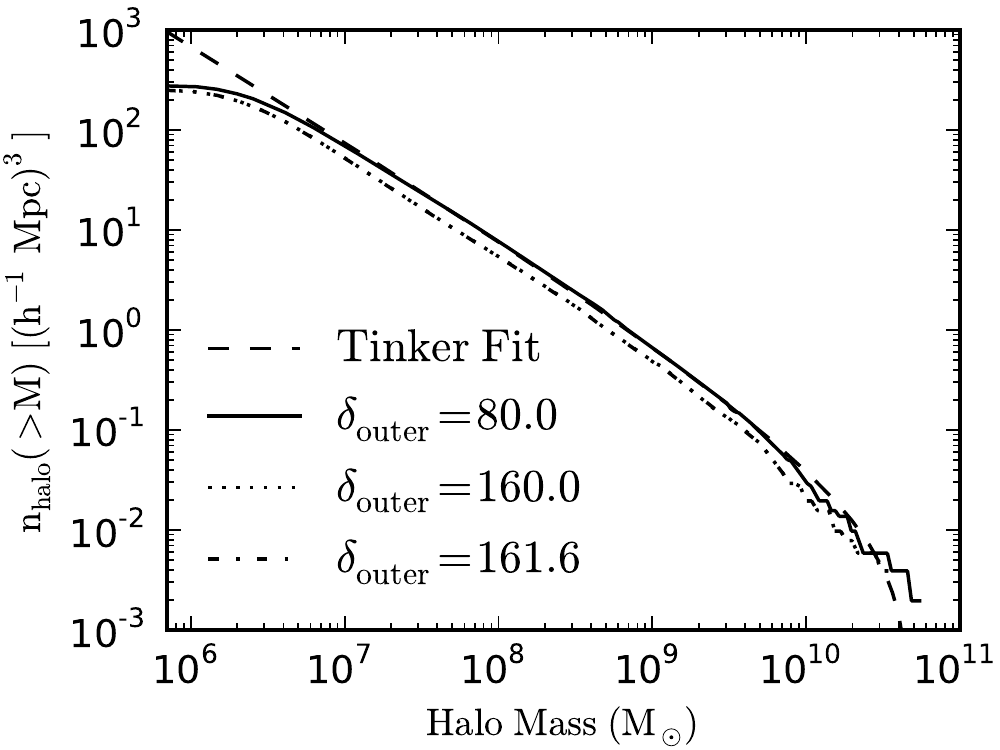}
   \caption{The Halo Mass Function for D1024. Three runs of Parallel HOP with default settings and varying the value of \douter, are plotted in
   comparison to the \citet{Tinker:2008p964} fit.
   The \douter=80.0 line very closely matches the $\Delta=300$ Tinker fit in the range of $10^7$--$10^9$ M$_{\odot}$.
   The departure from the fit at the low-mass end is attributable to insufficient mass and spatial resolution,
   and at the high end from a lack of statistical sampling of massive halos limited by the cosmological
   volume of the simulation.
   For comparison, the \douter=160.0 and 161.6 lines are lower in halo number density,
   and are also indistinguishable from one another.}
   \label{fig:hmf}
\end{figure}

\begin{figure}[htbp] 
   \centering
   \epsscale{.7}
   \plotone{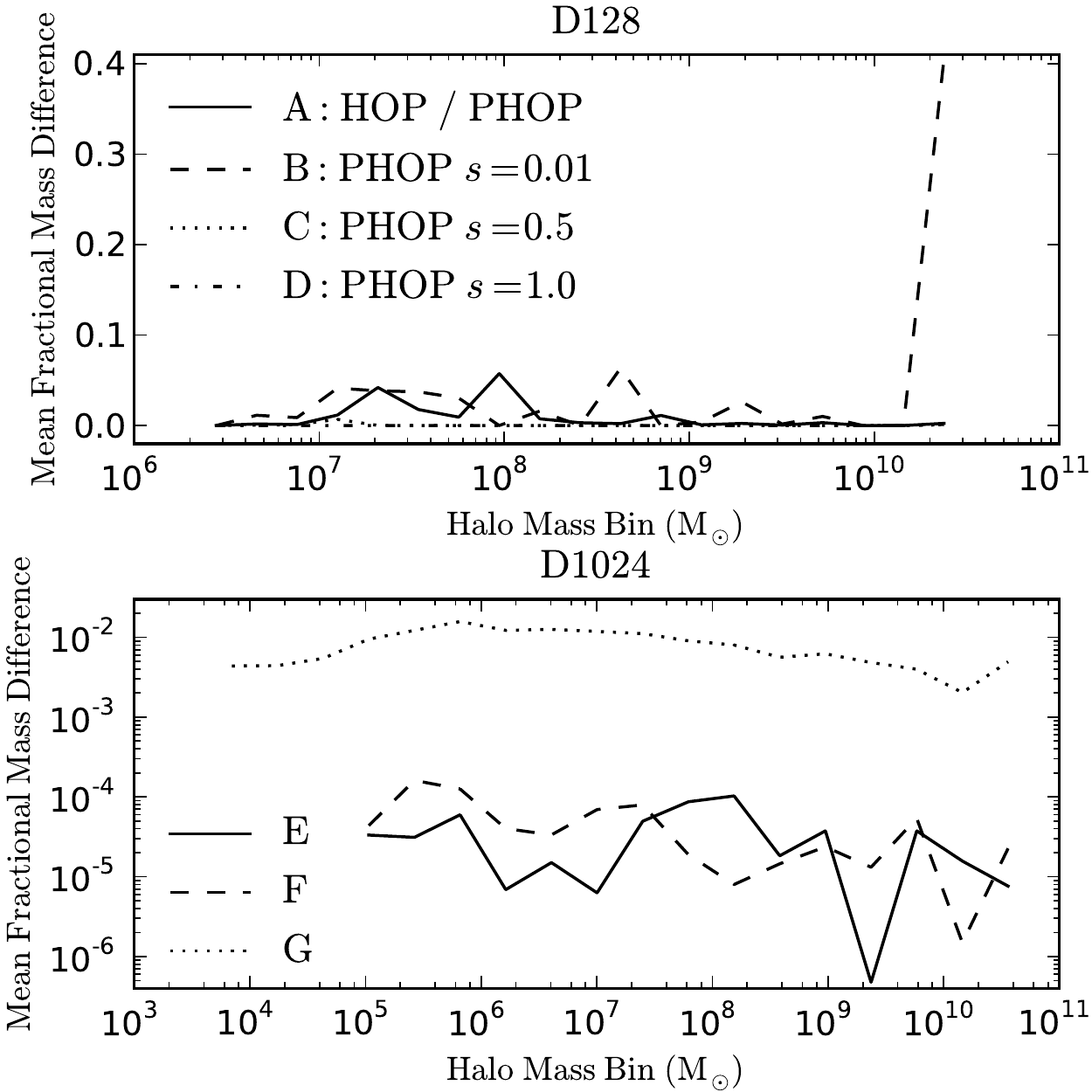}
   \caption{The mean absolute fractional change in halo mass in mass bins for cross-matched halos (by center of mass) between runs
   with different parameters. Fractional differences equal to zero mean the two runs are in agreement.
   Original HOP and Parallel HOP (PHOP) are run with the default settings (see Table \ref{table:parameters}) except noted.
   In the top plot are comparisons using D128, and using D1024 are in the bottom.
   The parameters varied for comparison are as follows.
   A: Original HOP compared to PHOP on one core with identical (\douter, \dsaddle, \dpeak, \Ndens, \Nmerge) values.
   B--D: Single-core PHOP and 8-core PHOP, varying the safety factor $s$.
   E: PHOP on 64 cores with \douter=160.0, with and without premerging.
   F: PHOP on 60 and 64 cores with \douter=160.0 and premerging.
   G: PHOP on 64 cores with premerging, comparing \douter=160.0 and 161.6.
   }
   \label{fig:params}
\end{figure}

\begin{figure}[htbp] 
   \centering
   \epsscale{0.65}
   \plotone{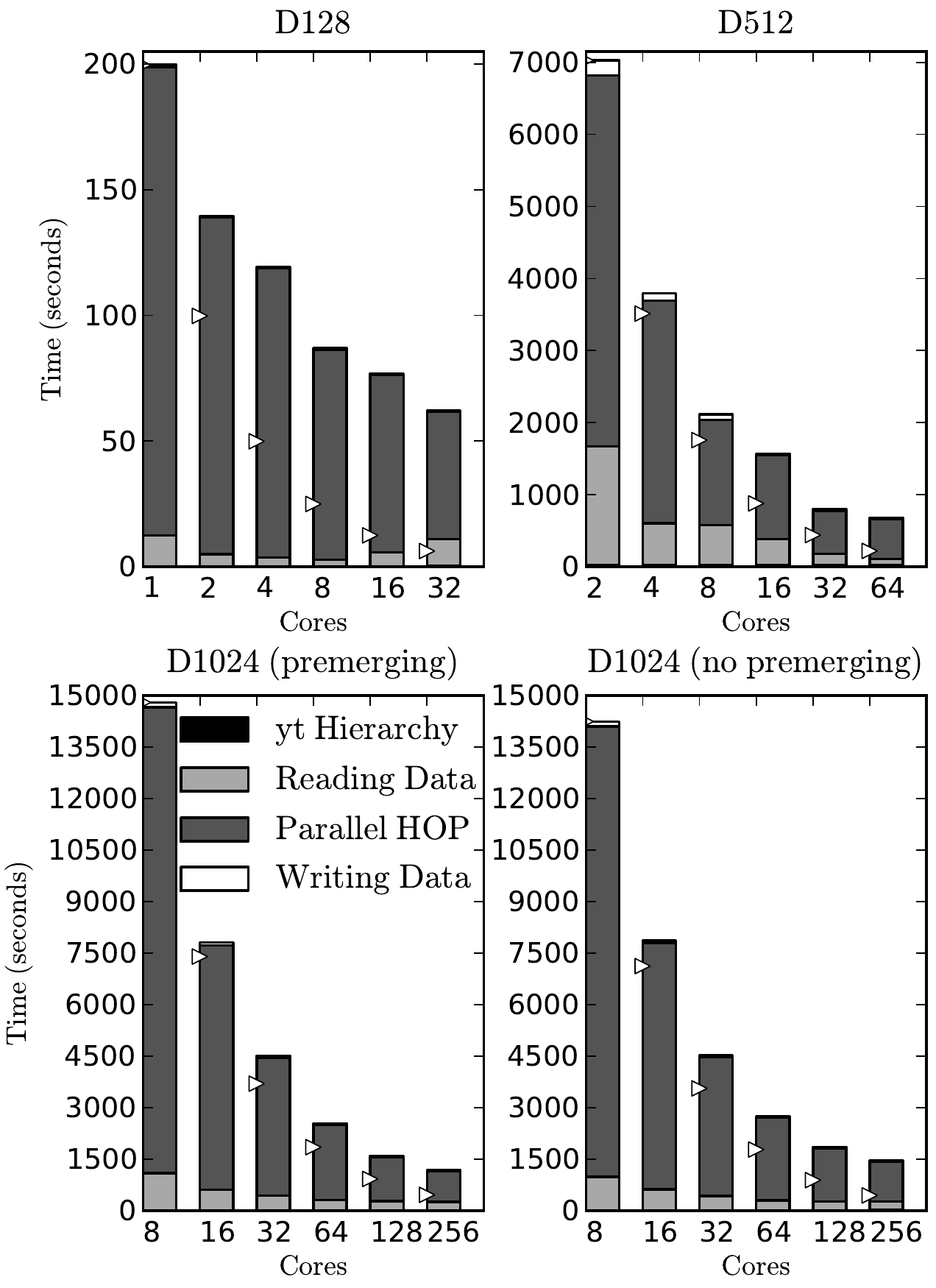}
   \caption{Full timings for Parallel HOP on three datasets. Each timing block shows the real-world `wallclock' time taken
   collectively for all tasks for each step. The triangles show linear scaling based on the smallest core count timing.}
   \label{fig:full_timings}
\end{figure}

\begin{figure}[htbp] 
   \centering
   \epsscale{0.65}
   \plotone{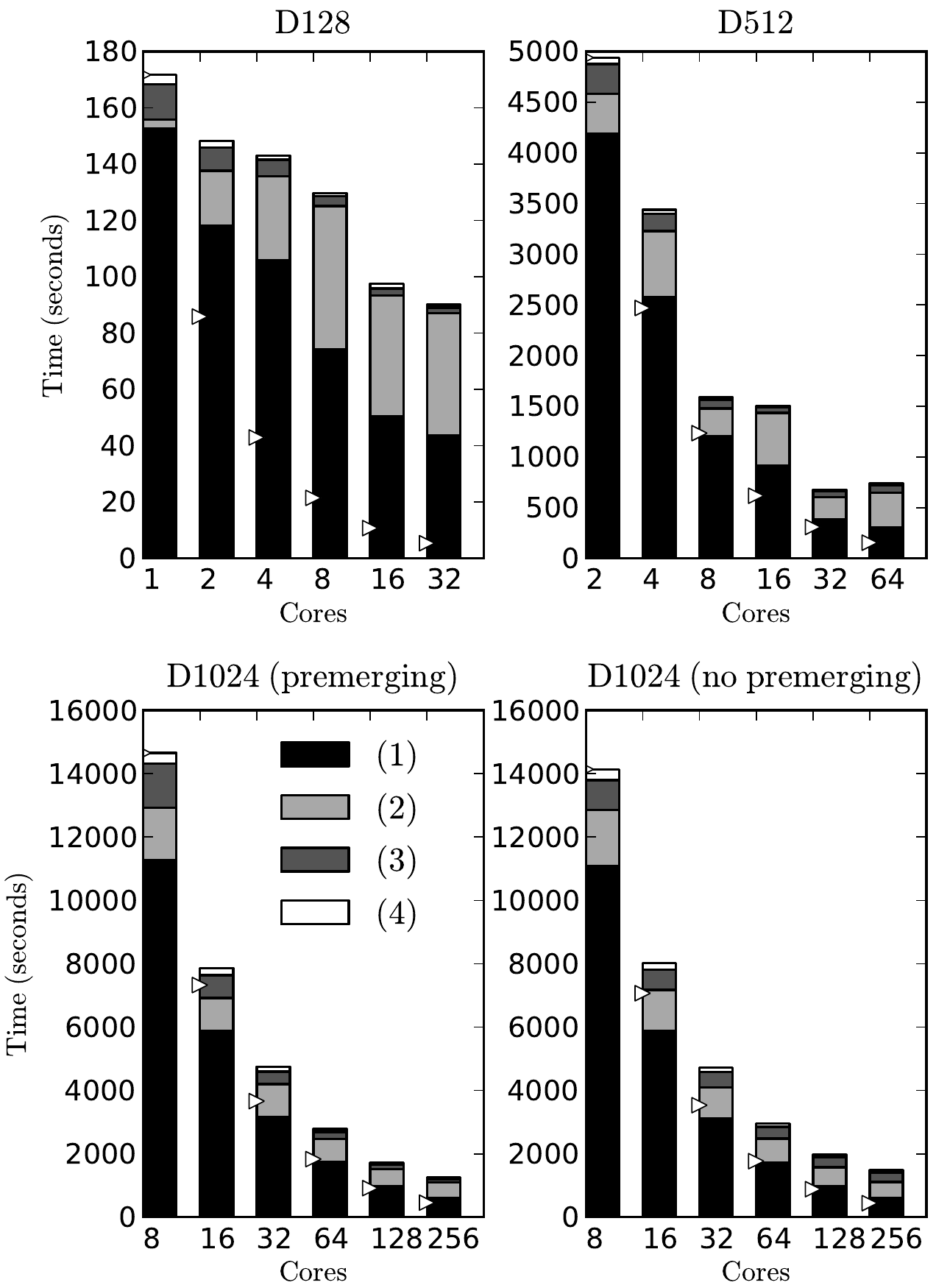}
   \caption{Timings for the `Parallel HOP' step. Each timing block is a sum of maximum time taken for several
   separate steps that are described by the timing block label: (1) - \kdtree\ Searching,
   (2) - MPI and related functions, (3) Halo Creation, and (4) Halo Statistics.
   The triangles show linear scaling based on the smallest core count timing.}
   \label{fig:pHOP_timings}
\end{figure}

\begin{figure}[htbp] 
   \centering
   \epsscale{0.65}
   \plotone{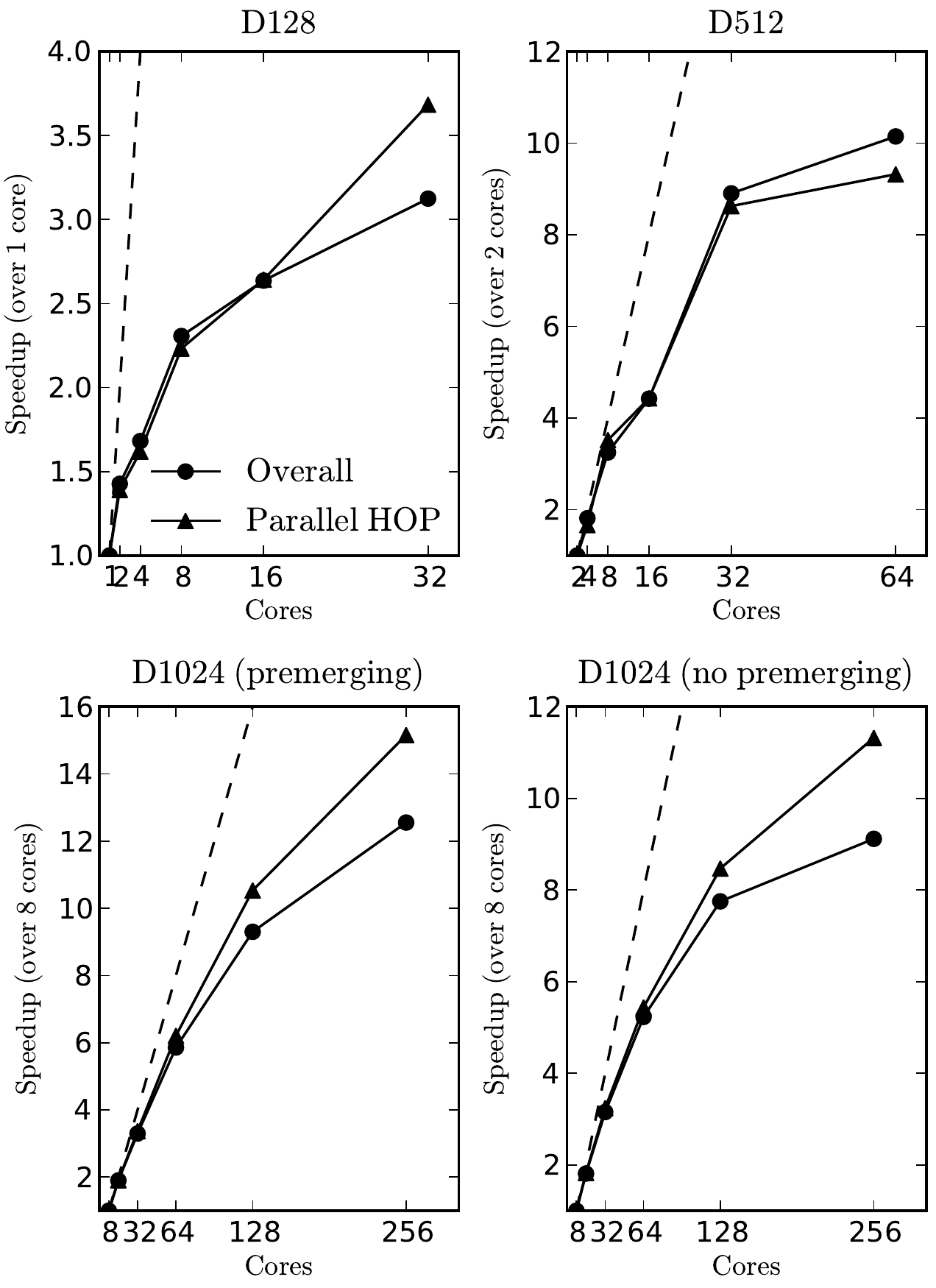}
   \caption{Overall and `Parallel HOP' only speedups. Note that the label for 16 cores is missing on the D1024 plots for legibility.
   The dashed lines show linear scaling based on the smallest core count timing.}
   \label{fig:pHOP_speedup}
\end{figure}

\begin{figure}[htbp] 
   \centering
   \epsscale{.4}
   \plotone{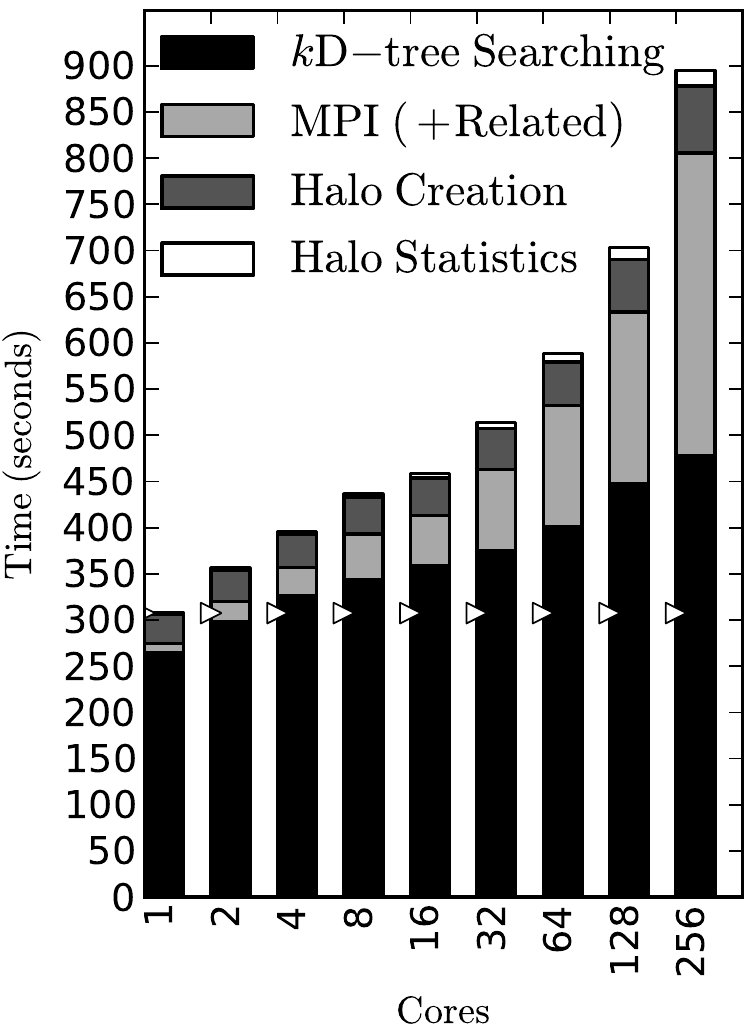}
   \caption{Weak-scaling timings on data randomly subsampled from D1024.
   The size of the dataset analyzed in each
   run is $N\times(1024^3 / 256 = 4,194,304)$, where $N$ is the number of cores.
   The rise in analysis time is a result of a nonlinear increase in workload with particle count.
   The default settings of Parallel HOP are used for the timings.
   The triangles show linear scaling based on the single-core timing.}
   \label{fig:pHOP_weak}
\end{figure}

Referring to the numbered glyphs in Figure \ref{fig:logicflow}, below is a list that gives the names
of source files and functions that accomplish each part of Parallel HOP.
Within the yt source, there are three files that contain Parallel HOP functions.
They are \texttt{yt/lagos/HaloFinding.py}, \texttt{yt/lagos/parallelHOP/parallelHOP.py}, and
\texttt{yt/extensions/kdtree/fKD.f90}, which are referred to as \HF, \PH\ and \KD, respectively.

\begin{enumerate}

\item The command invoked by a user to run Parallel HOP is \texttt{parallelHF}, which is contained in \HF.
It sets default initial values
and returns the list of halo objects from Parallel HOP at the very end of the analysis. In particular, depending on
the user's settings, the subvolumes are defined in this function (see below), and the padding lengths
are calculated, but not filled with particles until later.

\item If load-balancing is set, \HF: \texttt{\_subsample\_points}
subsamples the particles, and passes the subsampled points to \HF: \texttt{\_recursive\_divide} which
defines the load-balanced subvolumes.

\item The particle data is read off disk in parallel in \HF: \texttt{parallelHOPHaloList}. This function then
calls the function \PH: \texttt{\_chain\_hop}, which is the entry-point function into the
halo finding part of Parallel HOP.

\item Using the subvolume boundaries and padding lengths calculated earlier,
each task finds its set of neighbors and their padding lengths in \PH: \texttt{\_global\_padding} and \PH: \texttt{\_global\_bounds\_neighbors}.

\item Based on the values of padding for neighboring subvolumes,
\PH: \texttt{\_is\_inside} marks particles to be communicated to neighbors in \PH: \texttt{\_communicate\_padding\_data}
(and subsequent communication steps),
which fills the padding with particles on all tasks.

\item The particle data must be prepared prior to calling the Fortran \kdtree\ in \PH: \texttt{\_init\_kd\_tree}.
This then calls \KD: \texttt{create\_tree}
which builds the \kdtree\ and the resorted particle data for search optimization (if set by the user).
Next, \KD: \texttt{chainHOP\_tags\_dens} is called to calculate the density for all the particles
locally on each task, including the padding.
\PH: \texttt{\_densestNN} finds the densest nearest neighbor for each particle,
which then is used to build the initial chains of particles in \PH: \texttt{\_build\_chains} on each task independently.

\item If set, proto-halos are premerged on all tasks independently in \PH: \texttt{\_preconnect\_chains}.

\item In order to reconnect chains across subvolume boundaries, the chains must have globally unique identifiers,
which is set in \PH: \texttt{\_globally\_assign\_chainIDs}.
Also, because links across boundaries only
go `uphill', each task must have a lookup table of the densities of all the initial chains,
which is distributed in \PH: \texttt{\_create\_global\_densest\_in\_chain}.
Once both of those are done, the chains can be reconnected in \PH: \texttt{\_connect\_chains\_across\_tasks}.

\item Prior to the final merging step, padded particles must be assigned to chains authoritatively using
communication in function \PH: \texttt{\_communicate\_annulus\_chainIDs}.

\item The global boundary density lookup table is first built locally on each task in \PH: \texttt{\_connect\_chains},
and then merged globally in \PH:\texttt{\_make\_global\_chain\_densest\_n}.

\item Processing of the global lookup table happens in \PH: \texttt{\_build\_groups}, which defines the
final set of particle halos. Using the list of halos, particles are assigned to their halo in \PH: \texttt{\_translate\_groupIDs}.
Halo total sizes and masses, and centers of masses are calculated in \PH: \texttt{\_precompute\_group\_info}.

\end{enumerate}

Next, the particle velocities are read in, and the bulk velocity of halos calculated (\HF: \texttt{parallelHOPHaloList}).
The halo objects are created and sorted by halo mass (\HF: \texttt{\_parse\_output} and \texttt{\_join\_halolists}).
The list of halo objects can then be written to a text file (\HF: \texttt{write\_out}),
and the halo particle data stored to HDF5 files (\HF: \texttt{write\_particle\_lists}).

\acknowledgments

This work was partially supported by NSF grants AST-0708960 and AST-0808184.
Computations were carried out on the Triton Resource of the San Diego Supercomputer
Center and on Kraken at the National Institute for Computational Sciences.

\clearpage

\begin{deluxetable}{rcccc}
\tablecolumns{5}
\tablewidth{0pc} 
\tablecaption{Cores Per Node}
\tablehead{
\colhead{} & \multicolumn{3}{c}{Runs} \\
\cline{2-5}\\
\colhead{} & \colhead{D128} & \colhead{D512} & \colhead{D1024} & \colhead{D1024 (weak-scaling)}\\
\colhead{Total Cores} & \colhead{\em TCC (8 cores/node)} & \colhead{\em TCC} & \colhead{\em PDAF (32 cores/node)} & \colhead{\em PDAF}}
\startdata
1 & 1 & \nodata & \nodata & 1 \\
2 & 2 & 1 & \nodata & 2 \\
4 & 4 & 1 & \nodata & 4 \\
8 & 8 & 2 & 1 & 8 \\
16 & 8 & 2 & 2 & 16\\
32 & 8 & 4 & 4 & 32 \\
64 & \nodata & 8 & 8 &32 \\
128 & \nodata & \nodata & 16 & 32 \\
256 & \nodata & \nodata & 32 &32 \\
\enddata
\label{table:corespernode}
\end{deluxetable} 

\begin{deluxetable}{ccc}
\tablecolumns{3}
\tablewidth{0pc} 
\tablecaption{Parallel HOP Parameters}
\tablehead{
\colhead{Parameter} & \colhead{User Controlled?} & \colhead{Default}
}
\startdata
\multicolumn{3}{c}{`Physical' Parameters, HOP and Parallel HOP} \\
\cline{1-3}\\
\douter: Outer Density Threshold & Yes & 80.0 \\
\dsaddle: Proto-Halo Merging Threshold & No & 2.5\douter \\
\dpeak: Proto-Halo Threshold & No & 3\douter \\
\Ndens: Smoothing Kernel Size & No & 65 \\
\Nmerge: Chain Boundary Search & No & 4 \\
\cline{1-3} \\
\multicolumn{3}{c}{`Technical' Parameters, Parallel HOP Only} \\
\cline{1-3} \\
Load-balancing & Yes & On \\
Load-balancing Subsample Fraction & Yes & 0.03 \\ 
Directional Padding & Yes & On \\
Padding Safety ($s$) & Yes & 1.5 \\
Chain Premerging & Yes & On \\
\kdtree\ Speedup & Yes & On
\enddata
\label{table:parameters}
\end{deluxetable} 

\clearpage

\bibliographystyle{astron}
\bibliography{refs}

\clearpage

\end{document}